\documentclass[sigconf]{acmart}

\def\BibTeX{{\rm B\kern-.05em{\sc i\kern-.025em b}\kern-.08emT\kern-.1667em\lower.7ex\hbox{E}\kern-.125emX}}

\copyrightyear{}
\acmYear{}
\setcopyright{acmlicensed}
\acmConference[]{}{}{}
\acmBooktitle{}
\acmPrice{}
\acmDOI{}
\acmISBN{}

\usepackage[ruled,vlined]{algorithm2e}
\usepackage{array}
\usepackage{amsfonts}
\usepackage{amsmath}
\usepackage{booktabs}
\usepackage{caption}
\usepackage{comment}
\usepackage{enumitem}
\usepackage{graphicx,epstopdf}
\usepackage{hyperref}
\usepackage{multirow}
\usepackage{natbib}
\usepackage{newtxmath}
\usepackage{subcaption}
\usepackage{threeparttable}

\graphicspath{ {figures/} }

\newcommand{\eg}{\emph{e.g.},\xspace}

\newcommand{\myparagraph}[1]{\vspace{0.5em}\noindent\textbf{#1}}
\newcommand{\ie}{\emph{i.e.,}\xspace}

\newcommand{\etal}{\emph{et~al.}\xspace}
\newcommand\FramedBox[3]{%
  \setlength\fboxsep{3pt}
  \fbox{\parbox[t][#1][c]{#2}{#3}}}
\newcolumntype{x}[1]{>{\raggedright\arraybackslash}p{#1}}

\SetCommentSty{mycommfont}

\makeatletter 
\g@addto@macro{\@algocf@init}{\SetKwInOut{Output}{Output}} 
\makeatother

\begin{document}

\title{Knowledge Enhanced Multi-modal Fake News Detection}

\author{Yi Han, Amila Silva, Ling Luo, Shanika Karunasekera, Christopher Leckie}
\email{{yi.han@, amila.silva@student., ling.luo@, karus@, caleckie@}unimelb.edu.au}
\affiliation{%
  \institution{School of Computing and Information Systems\\The University of Melbourne}
  \streetaddress{}
  \city{}
  \state{}
  \postcode{}
}

\begin{abstract}
Recent years have witnessed the significant damage caused by various types of fake news. Although considerable effort has been applied to address this issue and much progress has been made on detecting fake news, most existing approaches mainly rely on the textual content and/or social context, while knowledge-level information---entities extracted from the news content and the relations between them---is much less explored. Within the limited work on knowledge-based fake news detection, an external knowledge graph is often required, which may introduce additional problems: it is quite common for entities and relations, especially with respect to new concepts, to be missing in existing knowledge graphs, and both entity prediction and link prediction are open research questions themselves. Therefore, in this work, we investigate \textbf{knowledge-based fake news detection that does not require any external knowledge graph.} Specifically, our contributions include: (1) transforming the problem of detecting fake news into a subgraph classification task---entities and relations are extracted from each news item to form a single knowledge graph, where a news item is represented by a subgraph. Then a graph neural network (GNN) model is trained to classify each subgraph/news item. (2) Further improving the performance of this model through a simple but effective multi-modal technique that combines extracted knowledge, textual content and social context. Experiments on multiple datasets with thousands of labelled news items demonstrate that our knowledge-based algorithm outperforms existing counterpart methods, and its performance can be further boosted by the multi-modal approach.
\end{abstract}

%
%

%

\maketitle

\section{Introduction}\label{sec:intro}
The prevalence of fake news\footnote{Here we use the definition in~\cite{zhou_fake_2018}: \textit{fake news is intentionally and verifiably false news published by a news outlet}.} on social media has serious repercussions on our society. Especially when equipped with big data analysis, it can accurately reach specific target audiences to spread fear, aggravate hatred, and cause riots and violence.
 
While significant efforts have been made on fake news detection, most existing work focuses on the \textit{textual content} and \textit{social context}, where \textit{textual content} includes the news headline and body, and \textit{social context} means user interactions over social media. For example, once a news item is published online, it may be tweeted by multiple users. Each of these tweets and its retweets form a separate cascade~\cite{vosoughi_spread_2018}, and all the cascades form the propagation pattern/network (we use these two terms interchangeably in this work) of a news item.

Knowledge-level information, on the other hand, has been much less explored. Here knowledge refers to the entities in the news content and the relations between them. Typically, it can be represented in the format of a triple: (Subject, Predicate, Object), \ie SPO triple~\cite{noauthor_resource_2017}. Within the limited work on knowledge-based fake news detection, an external knowledge graph is often required. For example, Cui \etal~\cite{cui_deterrent_2020} incorporate an article-entity bipartite graph and a medical knowledge graph to better capture the important entities and guide the embedding of news articles to detect healthcare misinformation. However, it is unlikely for any knowledge graph to contain all possible entities and relations, and it is not a trivial task to accurately predict missing ones. To overcome this issue, in this work, we investigate \textbf{knowledge-based fake news detection that does not require any external knowledge graph.}

\begin{figure*}[ht!]
\centering
\begin{subfigure}{.75\textwidth}
  \centering
  \includegraphics[width=\textwidth]{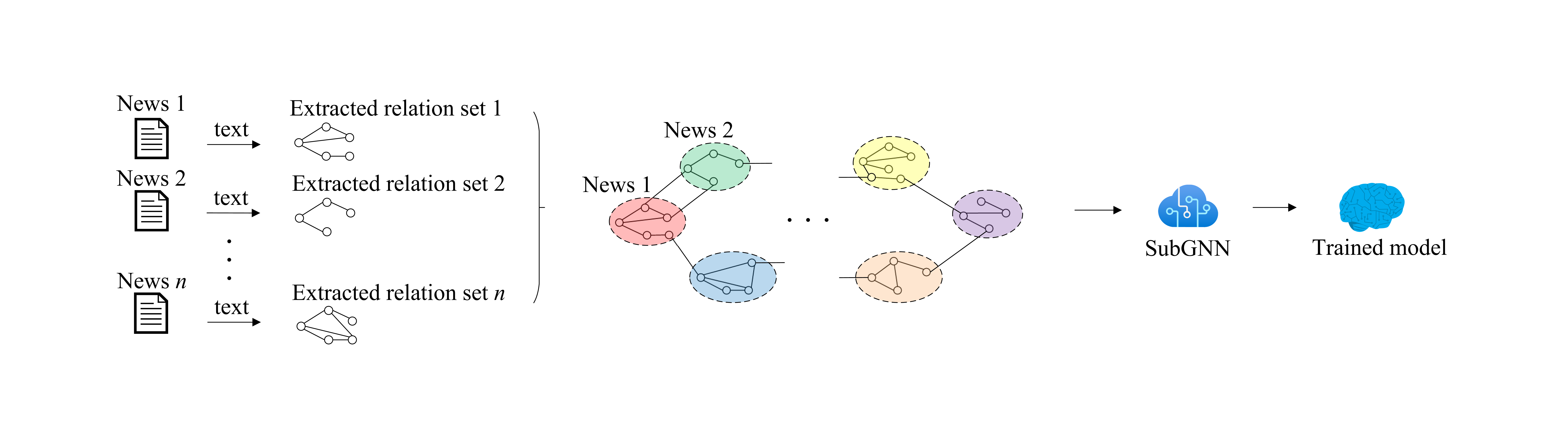}
  \caption{Knowledge-based fake news detection. Information is extracted from all news items to form a single knowledge graph, where each news item can be represented by a subgraph, so that the fake news detection problem becomes a subgraph classification task. A SubGNN model is then trained to classify each subgraph/news item.}
  \label{figure_overview_knowledge}
\end{subfigure}
\begin{subfigure}{0.75\textwidth}
  \centering
  \includegraphics[width=.9\textwidth]{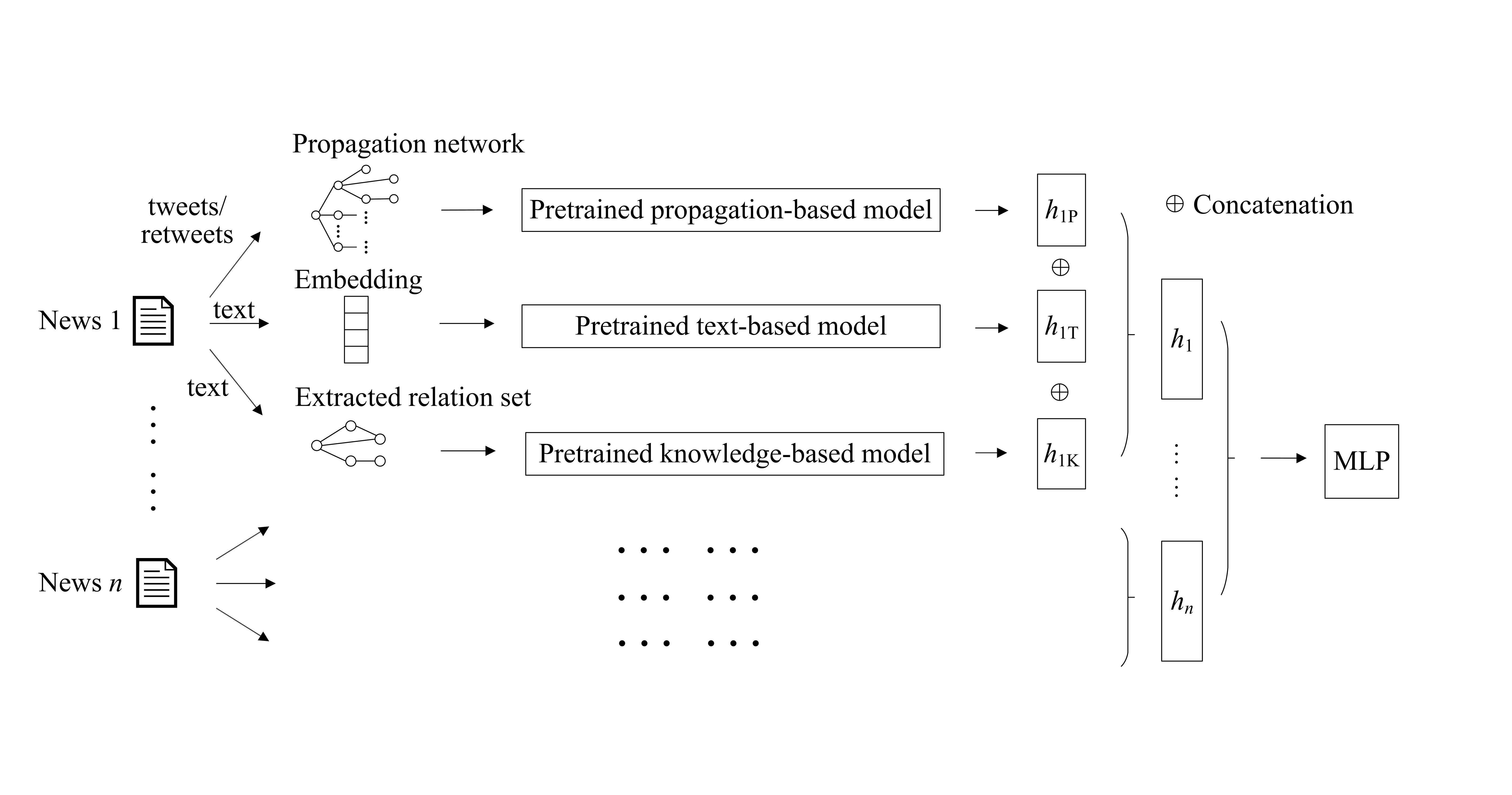}
  \caption{Multi-modal fake news detection that combines extracted knowledge, text content and propagation network. The propagation-, text- and knowledge-based models are pre-trained separately, and their generated embeddings are concatenated to train a MLP.}
  \label{figure_overview_multimodal}
\end{subfigure}
\caption{Overview of the proposed knowledge-based and multi-modal fake news detection algorithms.}
\label{figure_overview}
\end{figure*}

Pan \etal~\cite{pan_content_2018} approach this problem by constructing two knowledge graphs, with one on real news (\(KG_{T}\)) and one on fake news (\(KG_{F}\)). Each of these two knowledge graphs is then used to train a TransE~\cite{bordes_translating_2013} model, and a news item is verified by comparing the average or maximum bias of the two models over the extracted relations, where the bias of a triple \((s, p, o)\) is defined as \(||s+p-o||_{2}^{2}\). A news item is classified as ``Real" if the bias of the model trained on \(KG_{T}\) is smaller, and ``Fake" otherwise.

In our work, by contrast, we look at the problem from a different perspective: given the recent development on graph neural networks (GNNs) and their superior performance on non-Euclidean data, we investigate whether it is possible to transform fake news detection into a knowledge graph classification problem.

Initially, we test a method that constructs a separate knowledge graph for each news item, and then applies GNNs for graph-level classification. However, our preliminary experiments suggest that this method does not achieve satisfactory results, with an average accuracy only around 73\%. A key reason is that normally a very limited number of relations/triples can be extracted from each news item, which means that the constructed knowledge graph is too small to accurately verify the veracity of a news item.
    
Therefore, we propose a new approach and \textbf{transform the problem of fake news detection into a subgraph classification task}. Specifically, our contributions include:

\begin{itemize}[wide]
    \item Inspired by the work of~\cite{alsentzer_subgraph_2020}, which designs a GNN model called SubGNN for subgraph classification, we propose to extract entities and relations from each news item, all of which form a single knowledge graph where a news item is represented by a subgraph (Fig.~\ref{figure_overview_knowledge}), and then all subgraphs with their corresponding news labels (fake/real) are used to train a SubGNN model, so that the obtained model can classify each subgraph/news item. As demonstrated by the experimental results in Section~\ref{sec:experiment}, this method can achieve much better performance.
    
    \item To further improve the performance of the above model, we develop a simple but effective multi-modal fake news detection algorithm. In addition to extracted knowledge, other forms of information, such as the textual content and propagation network, can also contribute to detecting fake news. Specifically, as can be seen from Fig.~\ref{figure_overview_multimodal}, in our proposed method three models are first trained separately: (1) a propagation-based model~\cite{han_graph_2020} that verifies a news item purely on its propagation pattern; (2) a document classification model using hierarchical attention networks~\cite{yang_hierarchical_2016} that operates directly on the news content; and (3) a knowledge-based model proposed in this work. We demonstrate that by concatenating the generated embeddings of these three models to train a multilayer perceptron (MLP), this seemingly simple approach can outperform each individual model by a clear margin.
\end{itemize}

The remainder of this paper is organised as follows: 
Section~\ref{sec:problem} defines the research problem;
Sections~\ref{sec:kg} and~\ref{sec:subgraph} present our knowledge-based fake news detection algorithm which does not require any external knowledge graph; 
Section~\ref{sec:multimodal} introduces an architecture that combines knowledge-, text- and propagation-based models to facilitate multi-modal fake news detection; 
Section~\ref{sec:experiment} provides experimental results to demonstrate the effectiveness of our proposed methods; 
Section~\ref{sec:related} reviews previous work related to fake news detection on social media; 
and finally Section~\ref{sec:conc} concludes the paper and offers directions for future work.


\section{Problem Definition}\label{sec:problem}
Originally, the fake news detection problem can be defined as: given a set of labelled news items \(\mathcal{D} = \{\left(W_{i}, y_{i}\right) |\ i=1,\ 2,\ ...\}\), where \(W_{i} \in \mathcal{W}\) is the textual content for news item \(i\) (\ie a sequence of words), and \(y_{i} \in \mathcal{Y} = \{0\ \text{(Real)}, 1\ \text{(Fake)}\}\) is the label of \(W_{i}\), the goal is to learn a mapping \(g: \mathcal{W} \to \mathcal{Y}\) that classifies each news item.

In this work, we take a knowledge-based approach and break down the above formulation into the following two sub-problems:

\myparagraph{Knowledge graph construction.} The aim of the first step is to extract information from each news item to construct a knowledge graph. Conventionally, this involves named entity recognition and relation extraction, which have been extensively studied. However, we have tested several existing named entity-based relation extraction techniques, and our results suggest that when applied to news items, these methods normally generate a relatively small number of relations, which may lead to substantial information loss. Therefore, in the first sub-problem we design a new relation extraction algorithm that extracts a set of relations/triples \(R_{i} = \{(e_{ij}, r_{ij}, e_{ij}^{\prime}) | j = 1, 2, ...\}\) from news content \(W_{i}\), where \(e_{ij}, r_{ij}, e_{ij}^{\prime}\) contain one or multiple words in \(W_{i}\). Each of these triples means that \(e_{ij}\) and \(e_{ij}^{\prime}\) has the relation of \(r_{ij}\). For example, a triple (David Warner, troll, Virat Kohli) can be extracted from the sentence ``David Warner trolls Virat Kohli on Instagram over his grey beard" (Fig.~\ref{fig:example1}). Note that here we do not pre-define any relation type or named entity.

\myparagraph{Subgraph classification.} Once all the relations are extracted from news items, and a single knowledge graph is constructed (in the case where a part of the graph is isolated from the rest, we only keep the largest connected component), each news item is assigned to a subgraph based on its extracted relations. Therefore, the original fake news detection problem is transformed into a subgraph classification task formulated as follows: given a set of labelled sub-knowledge graphs \(\{\left(SG_{i}, y_{i}\right) |\ i=1,\ 2,\ ...\}\), where \(SG_{i} \in \mathcal{SG}\) represents the sub-knowledge graph that corresponds to news item \(i\), and \(y_{i} \in \mathcal{Y} = \{0\ \text{(Real)}, 1\ \text{(Fake)}\}\) is the label of \(SG_{i}\), then the goal is to learn a classifier \(f: \mathcal{SG} \to \mathcal{Y}\) that labels each subgraph. Note that here different subgraphs are not necessarily mutually exclusive and may contain common nodes.

In the following two sections, we explain in detail our solution to the above two problems.

\section{Knowledge Graph Construction}\label{sec:kg}
In order to extract relations from news items to build a knowledge graph, we have first tested the following options: (1) an OpenNRE~\cite{han_opennre_2019} model trained on the Wiki80 dataset\footnote{https://github.com/thunlp/OpenNRE/blob/master/benchmark/download\_wiki80.sh} with a BERT encoder, (2) an OpenNRE model trained on the Wiki80 dataset with a CNN encoder, (3) an OpenNRE model trained on the TACRED dataset\footnote{https://nlp.stanford.edu/projects/tacred/}~\cite{zhang_position-aware_2017} with a BERT encoder, (4) an ATLOP~\cite{zhou_document-level_2021} model trained on the DocRED dataset~\cite{yao_docred_2019}. The first three models are for sentence-level relation extraction, while the last one is for document-level relation extraction---hence it can extract more accurate relations than the other three models in our case.

Our experimental results suggest that all these models extract a relatively small number of relations from each news item, which may lead to substantial information loss. In addition, it is also unlikely for all the extracted relations to form a single graph. A main reason is that the relation types are pre-defined in these three datasets: Wiki80/TACRED/DocRED contains 80/41/96 relations, which cannot cover all the scenarios. As a result, in many cases even though a certain type of relation does exist, ``\textit{no\_relation}" is returned instead.

\begin{algorithm}[t]
 \LinesNumbered
 \SetKwInput{Input}{Input}
 \SetKwInput{Output}{Output}
 \Input{The textual content $W=\{w_0, w_1, ...\}$ of a news record}
 \Output{The list of relations $L$}
 $L \leftarrow \emptyset$;\\
 $W \leftarrow coreference\_resolution(W)$;\\
 $\{D_0, D_1, ...\} \leftarrow dependency\_parsing(W)$;\\
 $\{p_0, p_1, ...\} \leftarrow pos\_tagging(W)$;\\
 $\{l_0, l_1, ...\} \leftarrow lemmatization(W)$;\\

\For{$i$ in $\{0,1,2,....,|W|-1\}$}{
    \If{$p_i$ is a verb}{
        $\{left\_nodes , right\_nodes\} \leftarrow \{\emptyset, \emptyset\}$;\\ 
        $edge\_type \leftarrow l_i$;\\
        \For{$(w_k, r, w_i)$ in $D_i$}{
            \If{$l_k$ is ``not"}{$edge\_type \leftarrow ``not\_" + edge\_type$;}
            \ElseIf{$r$ in $\{nsubj\}$}{$left\_nodes \leftarrow left\_nodes \cup \{l_k\}$;}
            \ElseIf{$r$ in $\{dobj, iobj, attr, xcomp\}$}{$right\_nodes \leftarrow right\_nodes \cup \{l_k\}$;}
        }
        \For{$left\_node$ in $left\_nodes$}{
            \For{$right\_node$ in $right\_nodes$}{
                $L \leftarrow L \cup {(left\_node, edge\_type, right\_node)}$;\\
            }
        }
    }
}
 $return\text{ }L$ 
\caption{Relation Extraction}
\label{algo:relation_extraction}
\end{algorithm}

\begin{figure*}[t!]
    \centering
    \includegraphics[width=.6\textwidth]{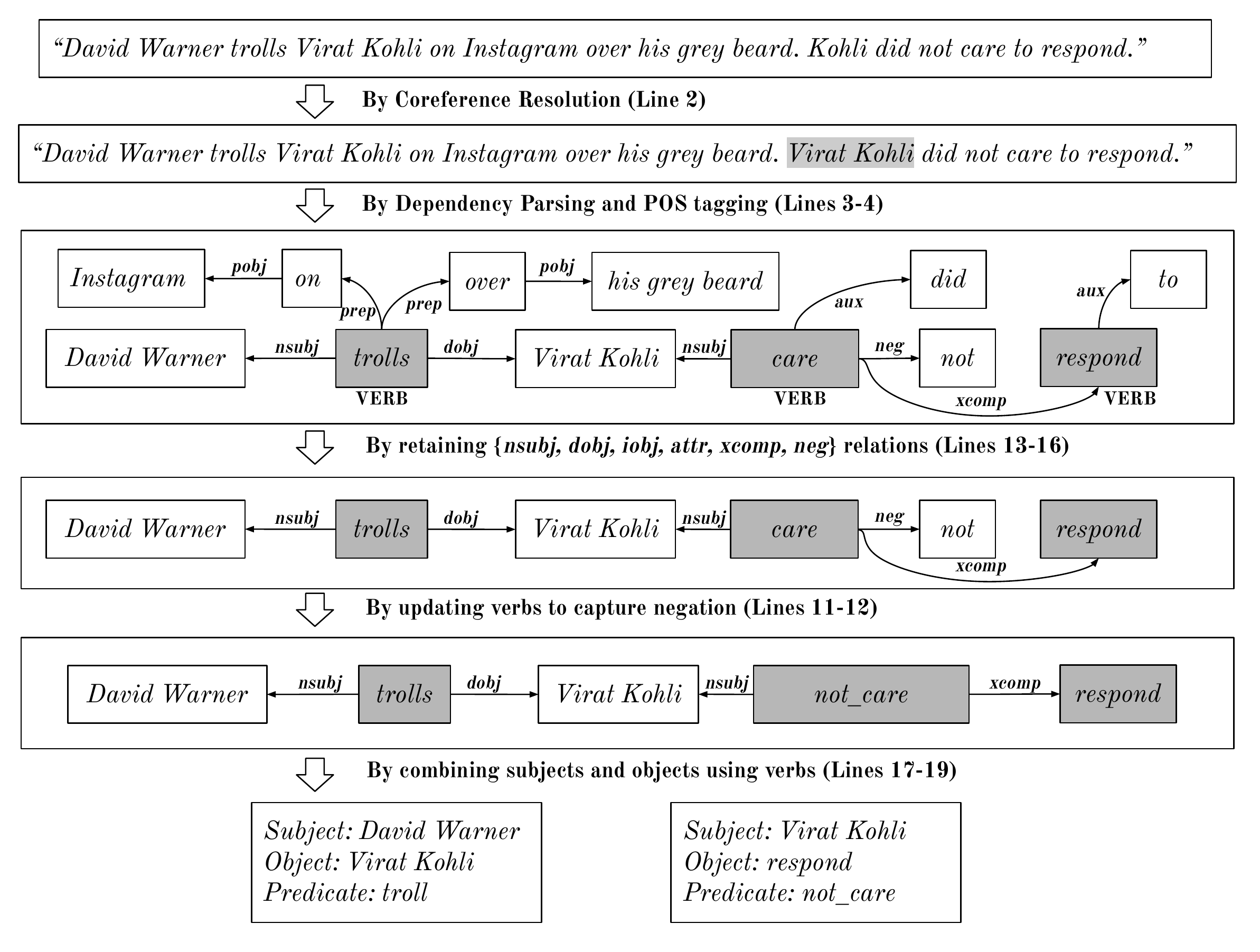}
    \caption{An illustration of Algorithm~\ref{algo:relation_extraction}.}
    \label{fig:example1}
\end{figure*}

One option is to create our own dataset from the collected news items with a larger number of pre-defined relations, and then retrain an OpenNRE or ATLOP model. However, in order to significantly increase the number of pre-defined relations, we also need a considerably larger number of relation instances for training. Considering that DocRED already has a total number of \(1,508,320\) instances, it is unlikely to obtain a much larger quantity from the thousands of collected news items.

Therefore, we design a new technique that can expand the number of extracted relations. Specifically, we do not pre-define any relation type, and instead focus on the verbs in each sentence, since verbs often play an important role in deciding the veracity of a statement. Algorithm~\ref{algo:relation_extraction} summarises our relation extraction method. As illustrated in Fig.~\ref{fig:example1}, for a news record, 

\begin{itemize}
    \item The title and the body text are concatenated to extract its textual content $W=\{w_0, w_1, ...\}$.
    \item Co-reference resolution\footnote{\href{https://spacy.io/universe/project/neuralcoref}{https://spacy.io/universe/project/neuralcoref}} is performed to replace all the mentions in $W$ that refer to the same real-world entity with a single token (Line 2). In the given example, ``Kohli" is replaced by ``Virat Kohli".
    \item The grammatical structure of each sentence in $W$ is extracted using the dependency parser available in Spacy\footnote{\href{https://spacy.io/usage/linguistic-features\#dependency-parse}{https://spacy.io/usage/linguistic-features\#dependency-parse}} (Line 3). This step returns a set of tuples $D_i$ for each word $w_i$. Each entry in $D_i$ is a tuple $(w_k, r, w_i)$, where $w_k$ is another word in the same sentence of $w_i$ that is related to word $w_i$ from the relation type $r$, \eg nominal subject ($nsubj$), direct object ($dobj$), open clausal complement ($xcomp$), as shown in the third box in Fig.~\ref{fig:example1}. Note that the relation here has different meaning from the relation between entities mentioned above in the problem definition.
    \item The Part-of-Speech (POS) tags $\{p_0, p_1, ...\}$ and the base-forms $\{l_0, l_1, ...\}$ of the words in $W$ are recovered using the POS tagger\footnote{\href{https://spacy.io/usage/linguistic-features\#pos-tagging}{https://spacy.io/usage/linguistic-features\#pos-tagging}} and the lemmatizer\footnote{\href{https://spacy.io/usage/linguistic-features\#lemmatization}{https://spacy.io/usage/linguistic-features\#lemmatization}} in Spacy (Line 4-5).
    \item The verbs in $W$ are identified by looping through POS tags of the words in $W$ (Line 6-7).
    \item For each identified verb $w_i$, the connected words in the dependency parse tree $D_i$ are analysed. If a negation, \ie words that reverse the meaning of a word, is found to be attached to $w_i$, $w_i$ is updated as ``\textit{not\_}" $+\  w_i$, \eg \textit{``care"} becomes \textit{``not\_care"} in the given example (Line 11-12). 
    \item Then, the connected nodes are categorised as either $left\_\allowbreak nodes$ or $right\_nodes$ based on their relation to $w_i$. If the relation type of a connected word is nominal subject ($nsubj$), it is added to $left\_nodes$ (Line 13-14). 
    \item Otherwise, if the relation to $w_i$ belongs to one of the following types: direct object ($dobj$), indirect object ($iobj$), attribute ($attr$), and open clausal complement ($xcomp$), it is added to $right\_nodes$ (Line 15-16). In the example of Fig.~\ref{fig:example1}, for the verb ``troll", ``David Warner"/``Virat Kohli" is categorised as $left\_nodes$/$right\_nodes$, respectively. In addition, even though ``\textit{respond}" is a verb in the second sentence, it will be extracted as a possible object of ``\textit{not\_care}" by our algorithm due to the \textit{xcomp} connection between them.
    \item In the end, the relation set is constructed by connecting each item in $left\_nodes$ with each item in $right\_nodes$ using the corresponding verb, \ie $w_i$ in the above example (Line 17-19).
\end{itemize}

\begin{figure*}[t!]
    \centering
    \includegraphics[width=.85\textwidth]{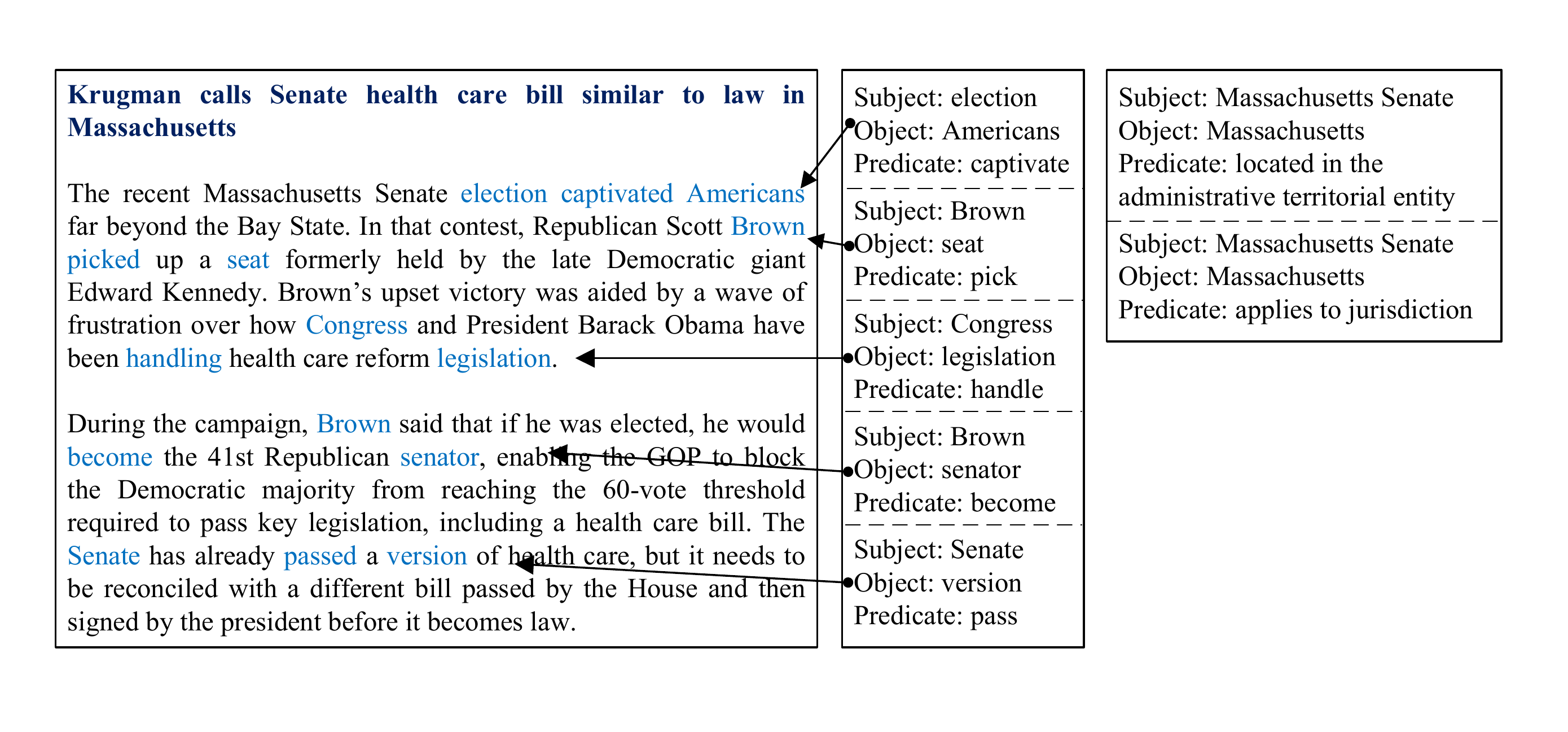}
    \caption{An example that compares the extracted relations using Algorithm~\ref{algo:relation_extraction} and by an ATLOP model. Left text box: the snippet of a news item. Middle text box: relations extracted by Algorithm~\ref{algo:relation_extraction}. Right text box: relations extracted by an ATLOP model trained on the dataset of DocRED.}
    \label{fig:example2}
\end{figure*}

Our experimental results show that by applying the above approach, substantially more (over 10 times) number of relations can be extracted from the news items in total. For example, as shown in Fig.~\ref{fig:example2}, in the left text box is the snippet of a news item\footnote{https://www.politifact.com/factchecks/2010/feb/04/paul-krugman/krugman-calls-senate-health-care-bill-similar-law-/}, and the relations extracted by our method are listed in the middle text box. As a comparison, the relations extracted by the ATLOP model are given on the right.

We have compared the cases where the knowledge graph is built using the relations extracted by (1) Algorithm~\ref{algo:relation_extraction}, (2) an ATLOP model trained on the DocRED dataset (since DocRED is especially collected for document-level relation extraction), (3) both (1) and (2). Our experiment suggests that the best result is achieved when only the relations extracted by Algorithm~\ref{algo:relation_extraction} are used.

Note that we are not proposing a general method for relation extraction, since it is unlikely for the defined rules to work for all cases, and Algorithm~\ref{algo:relation_extraction} is only for our knowledge-based fake news detection technique. 

After all the relations are extracted from news items, we add them one by one to form a single knowledge graph. Then each news item can be represented by a subgraph, based on the relations extracted from it. Therefore, the original problem of detecting fake news now transforms into classifying each of the subgraphs, as explained in the next section.

\section{Subgraph Classification}\label{sec:subgraph}
In this section, we first give a brief introduction to graph neural networks and SubGNN, and then explain how a sub-knowledge graph is classified in our case.

\subsection{Background on Graph Neural Networks}\label{sec:background_gnn}
Given a graph \(\mathcal{G} = \allowbreak \mathcal{(V, E)}\), with vertex/node set \(\mathcal{V}\), edge set \(\mathcal{E}\), and node feature set \(\mathcal{X} \in \mathbb{R}^{|\mathcal{V}|\times d}\) (\ie each node has \(d\) features), many GNN models can be formulated as a \textit{message passing framework}~\cite{battaglia_relational_2018,wu_comprehensive_2019} in which information is propagated from one node to another along edges. 

During each message-passing iteration \(k\), the embedding for node \(v \in \mathcal{V}\) is updated according to the information aggregated from \(v\)'s neighbourhood \(\mathcal{N}(v)\), which can be expressed as follows:

\(h_{v}^{(k+1)} = UPDATE \left( h_{v}^{(k)},\ AGGREGATE \left( \left\{ h_{u}^{(k)}\ |\ u \in \mathcal{N}(v)\right\} \right) \right)\),\\
where \(h_{v}^{0} = x_{v} \in \mathcal{X}\). The update step is often omitted by adding self-loops to the input graph, and the node embedding becomes:

\(h_{v}^{(k+1)} = AGGREGATE \left( \left\{h_{u}^{(k)}\ |\ u \in \mathcal{N}(v) \cup \left\{v\right\}\right\} \right)\).

Take a basic GNN for example~\cite{hamilton_graph_2020}, the model can be defined as: \(H^{(k+1)} = \sigma\left(\left(A+I\right) H^{(k)}M^{(k+1)}\right)\), where (1) \(H\) is the matrix of node embeddings/representations; (2) \(\sigma\) is a non-linear activation function, \eg the rectified linear unit (ReLU) function; (3) \(A \in \{0, 1\}^{|\mathcal{V}|\times |\mathcal{V}|}\) is the adjacency matrix: \(A_{i,j}=1\) if there is an edge from node \(i\) to node \(j\), and \(A_{i,j}=0\) otherwise; and (4) \(M\) is the weight matrix to be learned.

\subsection{Background on SubGNN}\label{sec:background_subgnn}
Most existing work on GNN focuses on node- and graph-level prediction tasks, while subgraphs are much less studied. To bridge this research gap, Alsentzer \etal~\cite{alsentzer_subgraph_2020} propose SubGNN that learns an embedding function \(\mathcal{F}: \mathcal{SG} \to \mathbb{R}^{d_{S}}\) to map each subgraph into a lower-dimensional representation.

Specifically, the embedding function \(\mathcal{F}\) captures features from three channels---\textit{position}, \textit{neighbourhood} and \textit{structure}, each of which has two subchannels---\textit{internal} and \textit{border}:

\begin{itemize}
    \item \textit{Position}. The \textit{internal position} of subgraph \(SG_{i}\) is defined as the distance between \(SG_{i}\)'s components---\(SG_{i}\) may contain a single connected component or multiple isolated components. The \textit{border position} is defined as the distance between \(SG_{i}\) and rest of \(\mathcal{G}\).
    \item \textit{Neighborhood}. The \textit{internal neighborhood} and \textit{border neighborhood} are defined as the identity of \(SG_{i}\)'s internal nodes and border nodes, respectively, where border nodes refer to nodes within the \(k\)-hop neighborhood of internal nodes.
    \item \textit{Structure}. The \textit{internal structure} is defined as the internal connectivity of \(SG_{i}\), while the \textit{border structure} is defined by edges connecting \(SG_{i}\)'s internal nodes to the border neighborhood.
\end{itemize}

\subsection{Classify Sub-Knowledge Graphs}
Each of the above channels can be used separately or together. We have tested multiple SubGNN models with different combinations of the three channels, and find that models that rely only on the structure channel gives the best result. This is because in our case, the internal and border structures, \ie how nodes are connected with each other within a subgraph and with the rest of the graph, are more informative for determining whether a news item is fake or not. Interestingly, the neighborhood channel does not perform as well as the other two, but this is consistent with the ablation study in the original work~\cite{alsentzer_subgraph_2020}.

In order to facilitate subgraph-level message passing for the structure channel, a number of connected components (each subgraph may contain single or multiple components) are randomly sampled via triangular random walks~\cite{boldi_arc-community_2012}---they are called anchor patches \(\mathcal{A} = \{A_{i}|i=1, 2, ..., n_{A}\}\). The embedding of a subgraph component (\(SC\)) is then represented in the following form:\\
\(h^{(k+1)}_{SC} = \sigma\left(M^{(k+1)} \cdot \left[h^{(k)}_{SC}; AGG \left( \left\{\gamma \left(SC, A_{i}\right) \cdot a_{i}|i=1, ..., n_{A} \right\} \right) \right] \right) \),\\
where \(\gamma\) is a pre-defined similarity function, \(a_{i}\) is the learned representation of \(A_{i}\), \(AGG\) is an aggregation function, \eg the sum operator, and \(M\) is a layer-wise trainable weight matrix.

If a subgraph contains multiple isolated connected components, its embedding is generated by concatenating the embeddings of all components.

\section{Knowledge Enhanced Multi-modal Fake News Detection}\label{sec:multimodal}
In addition to the extracted knowledge, the textual content itself and how the news item propagates through the social network also provide valuable information for detecting fake news. Therefore, in this section we study how to combine our knowledge-based approach with existing content- and propagation-based methods for more accurate detection.

Formally, given a set of labelled news items \(\{(SG_{i}, W_{i}, P_{i}, y_{i})\ |\ i=1,\ 2,\ ...\}\), where \(SG_{i} \in \mathcal{SG}\) is the sub-knowledge graph, \(W_{i} \in \mathcal{W}\) is the textual content, \(P_{i} \in \mathcal{P}\) is the propagation network, and \(y_{i} \in \mathcal{Y} = \{0\ \text{(Real)}, 1\ \text{(Fake)}\}\) is the label of the corresponding news, the goal is to learn a classifier \(f: \mathcal{SG} \times \mathcal{W} \times \mathcal{P} \to \mathcal{Y}\) that can label each news item.

As shown in Fig.~\ref{figure_overview_multimodal}, we first train three knowledge-, text- and propagation-based models separately on the same training dataset. Then for each training instance, we feed its \(SG_{i}, W_{i}, \text{and}\ P_{i}\) into the obtained models to generate three separate embeddings \(h_{i_{K}},\ h_{i_{T}}, \allowbreak h_{i_{P}}\), all of which are concatenated to form the final embedding \(h_{i} = h_{i_{K}}\oplus h_{i_{T}}\oplus h_{i_{P}}\). In the end, an MLP is trained on the set of embeddings \(\{h{_i}|i=1, 2, ...\}\).

Our experimental results in Section~\ref{sec:experiment} demonstrate that this seemingly simple method outperforms each individual model by a clear margin. Specifically, in this work we choose the following two as the text- and propagation-based models:

\begin{itemize}[wide]
    \item A document classification model using hierarchical attention networks~\cite{yang_hierarchical_2016}. The overall architecture of this model consists of four components: (1) a word encoder where words are embedded with a GRU-based sequence encoder; (2) a word-level attention layer where the importance of a word is measured by its similarity with a word-level context vector, which is jointly learned during the training process; (3) a sentence encoder that is also based on bidirectional GRU; and (4) a sentence-level attention layer that calculate the weight of a sentence in a similar way to (2). The reason why we choose this model is that previous work has shown that it performs better than other text-based models for fake news detection~\cite{shu_defend_2019}, such as LIWC~\cite{pennebaker_development_2015}, text-CNN~\cite{kim_convolutional_2014}.
    \item A propagation-based algorithm~\cite{han_graph_2020} that applies the GNN model of DiffPool~\cite{ying_hierarchical_2018} (built on top of GraphSage~\cite{hamilton_inductive_2017}) to verify a news item purely on its propagation pattern (as explained in the introduction) and the features of the corresponding Twitter users, including (1) whether the user is verified, (2) the timestamp when the user was created, (3) the number of followers, (4) the number of friends, (5) the number of lists, (6) the number of favourites; (7) the number of statuses, (8) the timestamp of the tweet. The adjacency matrix corresponding to the propagation pattern of a news item and the node feature matrix are fed as input for graph-level classification. We choose this model due to its simplicity and efficiency.
\end{itemize}

Note that (1) although the experimental results in the next section suggest that our knowledge-based model works well together with the above two models, especially the propagation-based model, they can be replaced by other options too; (2) in addition to text content and propagation network, there are other useful forms of information as well, including user replies, images and external knowledge graphs, especially domain-specific knowledge graphs. We leave more sophisticated fusion techniques for future work.

\section{Experimental Verification}\label{sec:experiment}
In this section, we empirically verify the effectiveness of our proposed knowledge-based and multi-modal fake news detection algorithms over two datasets with thousands of labelled news items.

\subsection{Datasets}
While there are a number of public datasets on fake news detection covering different domains, we choose the dataset of FakeNewsNet~\cite{shu_fakenewsnet_2018} in our work. FakeNewsNet contains labelled news from two websites: politifact.com\footnote{https://www.politifact.com/} and gossipcop.com\footnote{https://www.gossipcop.com/}---we call them PolitiFact and GossipCop hereafter. For each news item, the dataset provides both linguistic and visual information, all the tweets and retweets, as well as the information of the corresponding Twitter users. For more details please refer to~\cite{shu_fakenewsnet_2018}.

The reasons why we choose PolitiFact and GossipCop over other options include: (1) they align with the our definition of fake news---fake news is intentionally and verifiably false news published by a news outlet. Some public datasets on fake news detection, \eg Twitter16~\cite{ma_detect_2017}, Weibo~\cite{ma2016detecting}, Pheme~\cite{kochkina2017turing} are intended for detecting rumours, satires, misinformation, etc. (2) They provide accurate ground truth labels, which are collected using fact-checking websites. As a result, they are more accurate than those datasets where news items are weakly labelled by applying distant supervision techniques. For example, the datasets of ReCOVery~\cite{zhou2020recovery} and CoAID~\cite{cui2020coaid} label news records based on the reliability of the source. (3) They provide social context data, which is missing in datasets such as BuzzFeedNews~\cite{santia2018buzzface}, Ma-Twitter~\cite{ma2016detecting} and LIAR~\cite{wang_liar_2017}. This type of information is required by the chosen propagation-based approach to construct the propagation network for each news item.

\begin{table}[t!]
    \centering
    \caption{Sstatistics of the PolitiFact and GossipCop datasets.}
    \begin{tabular}{|c|c|c|}
    \hline
         Dataset & No. of Fake News & No. of Real News \\
         \hline
         PolitiFact & 185 & 225 \\
         \hline
         GossipCop & 4942 & 2520 \\
         \hline
    \end{tabular}
    \label{tab:dataset}
\end{table}

The statistics of the dataset are listed in Table~\ref{tab:dataset}. Note that the values are smaller than those reported in~\cite{shu_fakenewsnet_2018} because (1) many news items, especially fake news, have already been removed, (2) some tweets and retweets of a news items are no longer retrievable, and we cannot build the corresponding propagation pattern. Therefore, those news items are also excluded from our experiments.

\begin{table}[t!]
\caption{Performance comparison of fake news detection on the dataset of PolitiFact}
    \begin{threeparttable}
    \centering
    \begin{tabular}{c|c|c|c|c|c|c}
        \hline
        \ K\tnote{1} \ & \ T\tnote{1} \ & P\tnote{1} \ & Accuracy & Precision & Recall & F1\\
        \hline
        \checkmark & & & 0.837 & 0.844 & 0.801 & 0.836\\
        \hline
        & \checkmark & & 0.829	& 0.838	& 0.824	& 0.826\\
        \hline
        & & \checkmark & 0.854	& 0.855	& 0.852	& 0.852\\
        \hline
        \checkmark & \checkmark & & 0.858	& 0.853	& 0.842	& 0.857\\
        \hline
        & \checkmark & \checkmark & 0.884	& \textbf{0.922}	& 0.825	& 0.883\\
        \hline
        \checkmark & & \checkmark & 0.876	& 0.862	& 0.877	& 0.876\\
        \hline
        \checkmark & \checkmark & \checkmark & \textbf{0.919} & \textbf{0.913} & \textbf{0.913} & \textbf{0.919}\\
        \hline
        \hline
        \multicolumn{3}{c|}{Avg bias~\cite{pan_content_2018}} & 0.752 & 0.725 & 0.776 & 0.729\\
        \hline
        \multicolumn{3}{c|}{Max bias~\cite{pan_content_2018}} & 0.662 & 0.617 & 0.638 & 0.618\\
        \hline
    \end{tabular}
    \begin{tablenotes}
     \item[1] K: knowledge-based, T: text-based~\cite{yang_hierarchical_2016}, P: propagation-based~\cite{han_graph_2020}
   \end{tablenotes}
    \end{threeparttable}
    \label{tab:exp_plt}
\end{table}

\begin{table}[t!]
\caption{Performance comparison of fake news detection on the dataset of GossipCop}
    \begin{threeparttable}
    \centering
    \begin{tabular}{c|c|c|c|c|c|c}
        \hline
        \ K\tnote{1} \ & \ T\tnote{1} \ & P\tnote{1} \ & Accuracy & Precision & Recall & F1\\
        \hline
        \checkmark & & & 0.783 & 0.813 & 0.870 & 0.750\\
        \hline
        & \checkmark & & 0.788 & 0.767 & 0.747 & 0.754\\
        \hline
        & & \checkmark & \textbf{0.872} & 0.855 & 0.860 & \textbf{0.858}\\
        \hline
        \checkmark & \checkmark & & 0.800 & 0.823 & 0.888 & 0.768\\
        \hline
        & \checkmark & \checkmark & 0.861 & 0.882 & \textbf{0.910} & 0.843\\
        \hline
        \checkmark & & \checkmark & \textbf{0.879} & \textbf{0.916} & 0.899 & \textbf{0.866}\\
        \hline
        \checkmark & \checkmark & \checkmark & \textbf{0.871} & 0.903 & \textbf{0.900} & \textbf{0.856}\\
        \hline
        \hline
        \multicolumn{3}{c|}{Avg bias~\cite{pan_content_2018}} & 0.749 & 0.646 & 0.621 & 0.629\\
        \hline
        \multicolumn{3}{c|}{Max bias~\cite{pan_content_2018}} & 0.723 & 0.625 & 0.627 & 0.626\\
        \hline
    \end{tabular}
    \begin{tablenotes}
    \item[1] K: knowledge-based, T: text-based~\cite{yang_hierarchical_2016}, P: propagation-based~\cite{han_graph_2020}
    \end{tablenotes}
    \end{threeparttable}
    \label{tab:exp_gsp}
\end{table}

\subsection{Baselines}
Three baselines are considered: in addition to the two models chosen in Section~\ref{sec:multimodal}, \ie test-based~\cite{yang_hierarchical_2016} and propagation-based~\cite{han_graph_2020}, we also implement the knowledge-based method~\cite{pan_content_2018} described in the introduction.

In order to compare these models, the datasets are split as follows: 70\% of the data are used for training, 15\% are for validation, and the remaining 15\% are for test. In addition, all models are evaluated with the following commonly used metrics: accuracy, precision, recall and F1 score.

\myparagraph{Model hyper-parameters.} (1) For the text-based model, both the word embedding dimension and the GRU dimension are set to 100, the learning rate is 0.003, and the batch size is 32 for PolitiFact and 128 for GossipCop. (2) For the propagation-based model, the hyper-parameters for the DiffPool algorithm are: 2 pooling layers, 64 hidden dimensions and 64 embedding dimensions. In addition, the learning rate is 0.001, and the batch size is 128. (3) For our knowledge-based model, the hyper-parameters for SubGNN are selected from the following ranges: batch size \(\in [64, 128]\), learning rate \(\in [3e{\text -}5, \allowbreak 1e{\text -}3]\), number of layers \(\in [1, 4]\), number of structure anchor patches \(|A_{S}| \in [15, 45]\), and feed forward hidden dimension sizes \(\in [32, 64]\) with dropout \(\in [0.0, 0.4]\). (4) For our multi-modal approach, number of feed forward layers \(\in [2, 4]\) with hidden dimension sizes \(\in [8, 64]\) and dropout \(\in [0.0, 0.2]\). (5) For the baseline of~\cite{pan_content_2018}, the embedding dimensions of \(KG_{F}\) and \(KG_{T}\) are in the range of \([30, 50]\) for PolitiFact, and \([50, 100]\) for GossipCop.

\subsection{Performance Comparison of Fake News Detection}
Tables~\ref{tab:exp_plt} and \ref{tab:exp_gsp} show the performance comparison on the two datasets of PolitiFact and GossipCop (results in bold correspond to the best values or values less than \(0.01\) below the best values). The results suggest that:

\begin{itemize}
    \item Our knowledge-based model outperforms the baseline of~\cite{pan_content_2018} on both datasets.
    \item In terms of each single model of \(K,\ T\ \text{and}\ P\), the difference among their performances on the dataset of PolitiFact is not significant, but the propagation-based model achieves the best performance on the dataset of GossipCop, while the other two perform similarly again.
    \item In terms of the multi-modal approach, the performance of any combination of two models is almost always better than each individual model.
    \item If we compare the different combinations, \(K + T < T + P \approx K + P \leq K + T + P\): (1) since the knowledge- and text-based models rely on the same source of information---the news content, the combination of these two performs the worst. (2) When the knowledge- or text-based model is combined with the propagation-based model, they perform similarly and both better than (1). (3) The combination of all three models performs the best overall.
\end{itemize}
  
The above results suggest that while our knowledge-based approach that does not require any external knowledge graph is effective, combining it with other sources of information can further boost its performance.

\subsection{Time sensitivity analysis}
In addition to the metrics of accuracy, precision, recall and F1 score, it is important to understand how the performance of a model changes over time, since in real cases a deployed fake news detection system is likely to face highly dynamic and volatile data.

\begin{figure*}[t!]
\centering
\begin{subfigure}{\columnwidth}
  \centering
  \includegraphics[width=.95\columnwidth]{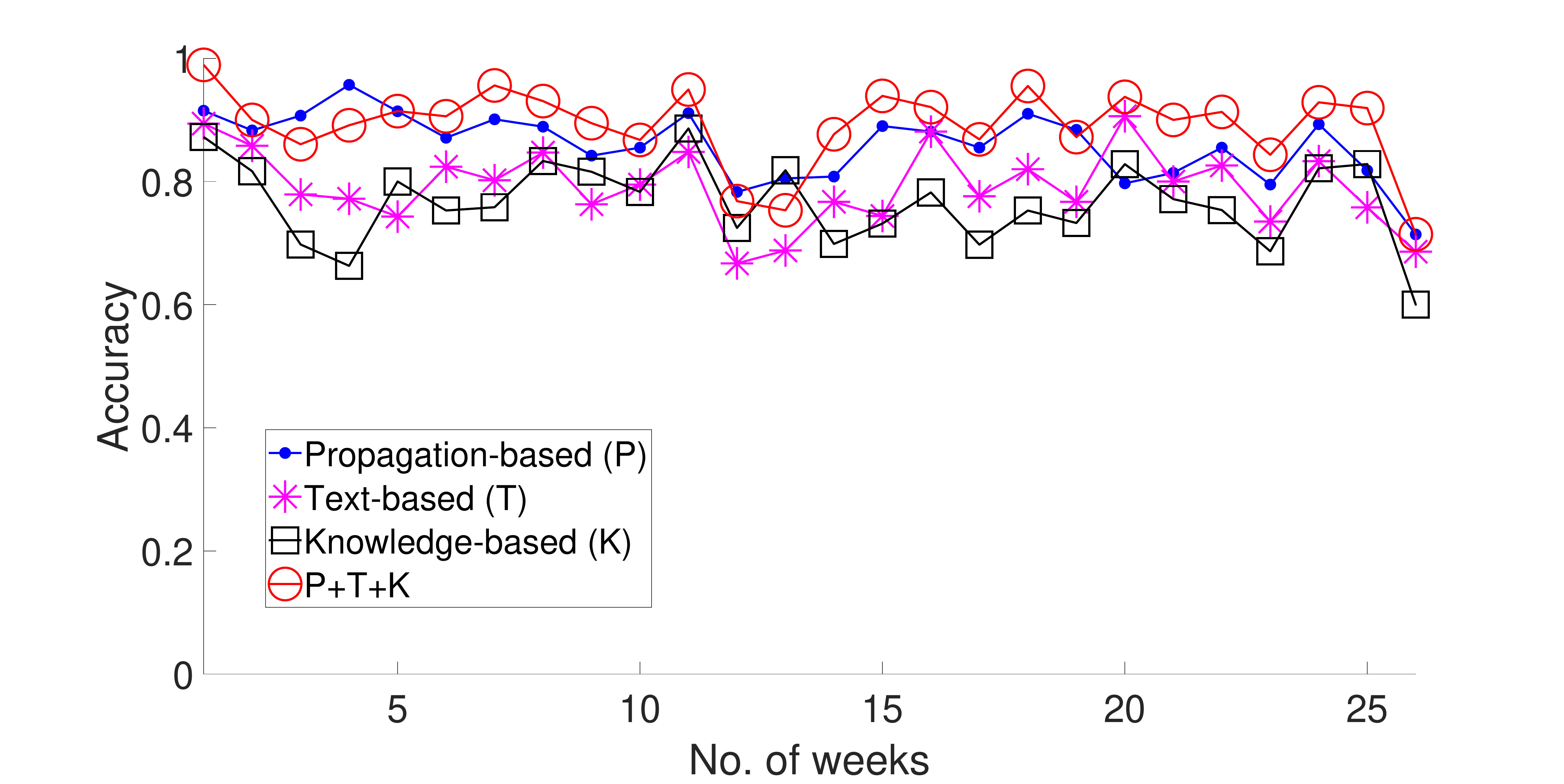}
  \caption{Accuracy}
  \label{figure_acc}
\end{subfigure}
\begin{subfigure}{\columnwidth}
  \centering
  \includegraphics[width=.95\columnwidth]{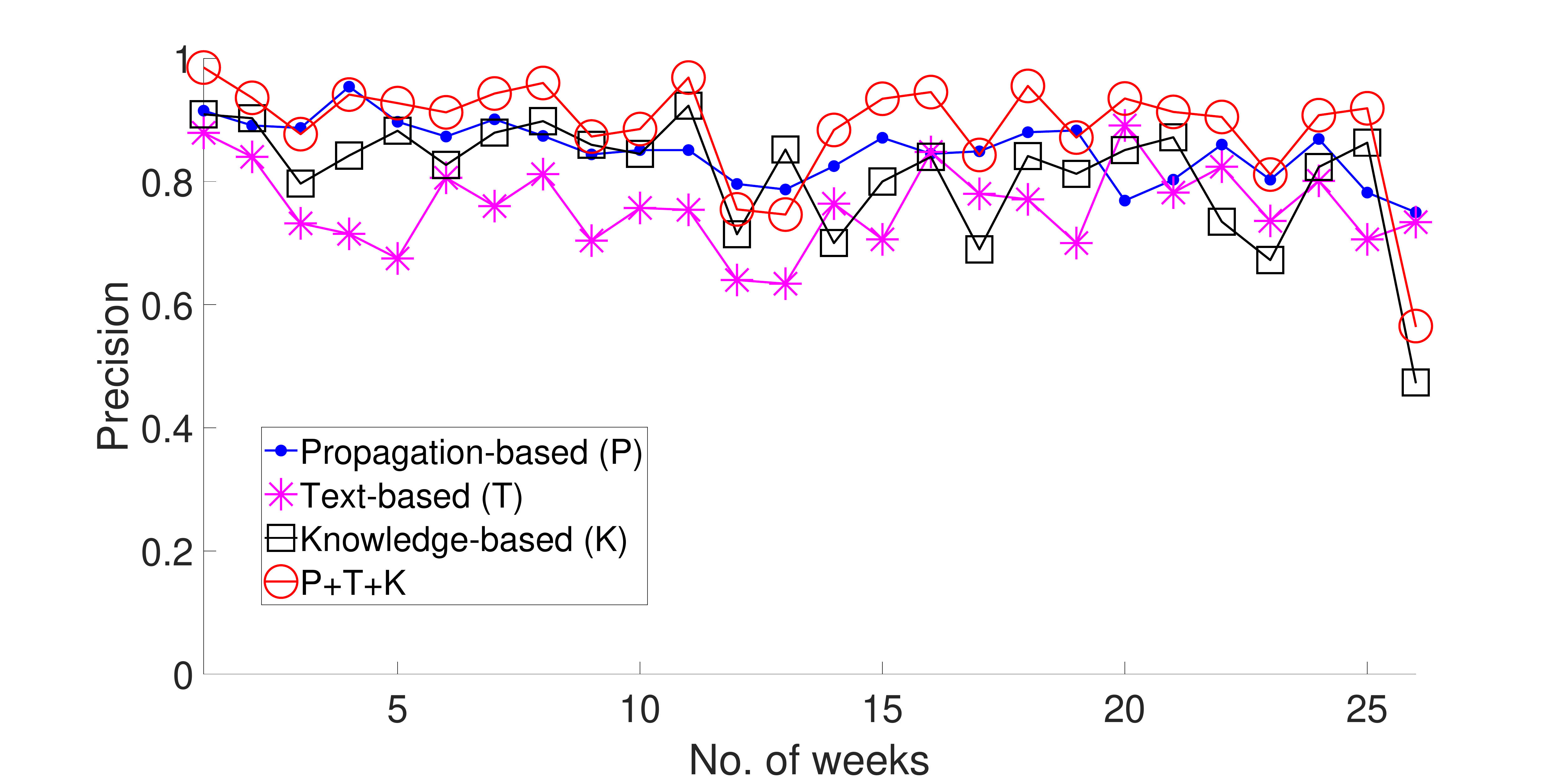}
  \caption{Precision}
  \label{figure_pre}
\end{subfigure}
\begin{subfigure}{\columnwidth}
  \centering
  \includegraphics[width=.95\columnwidth]{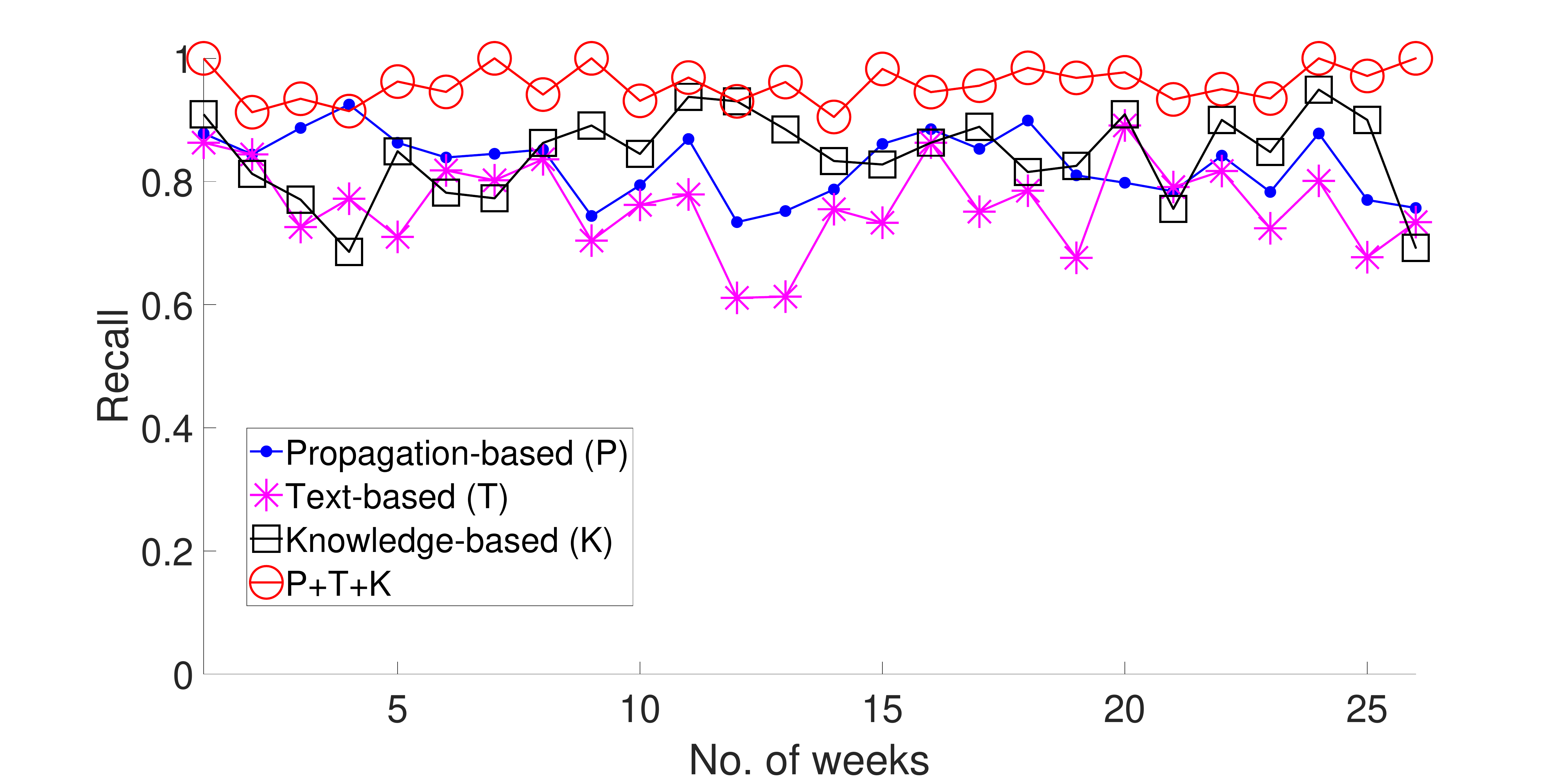}
  \caption{Recall}
  \label{figure_rec}
\end{subfigure}
\begin{subfigure}{\columnwidth}
  \centering
  \includegraphics[width=.95\columnwidth]{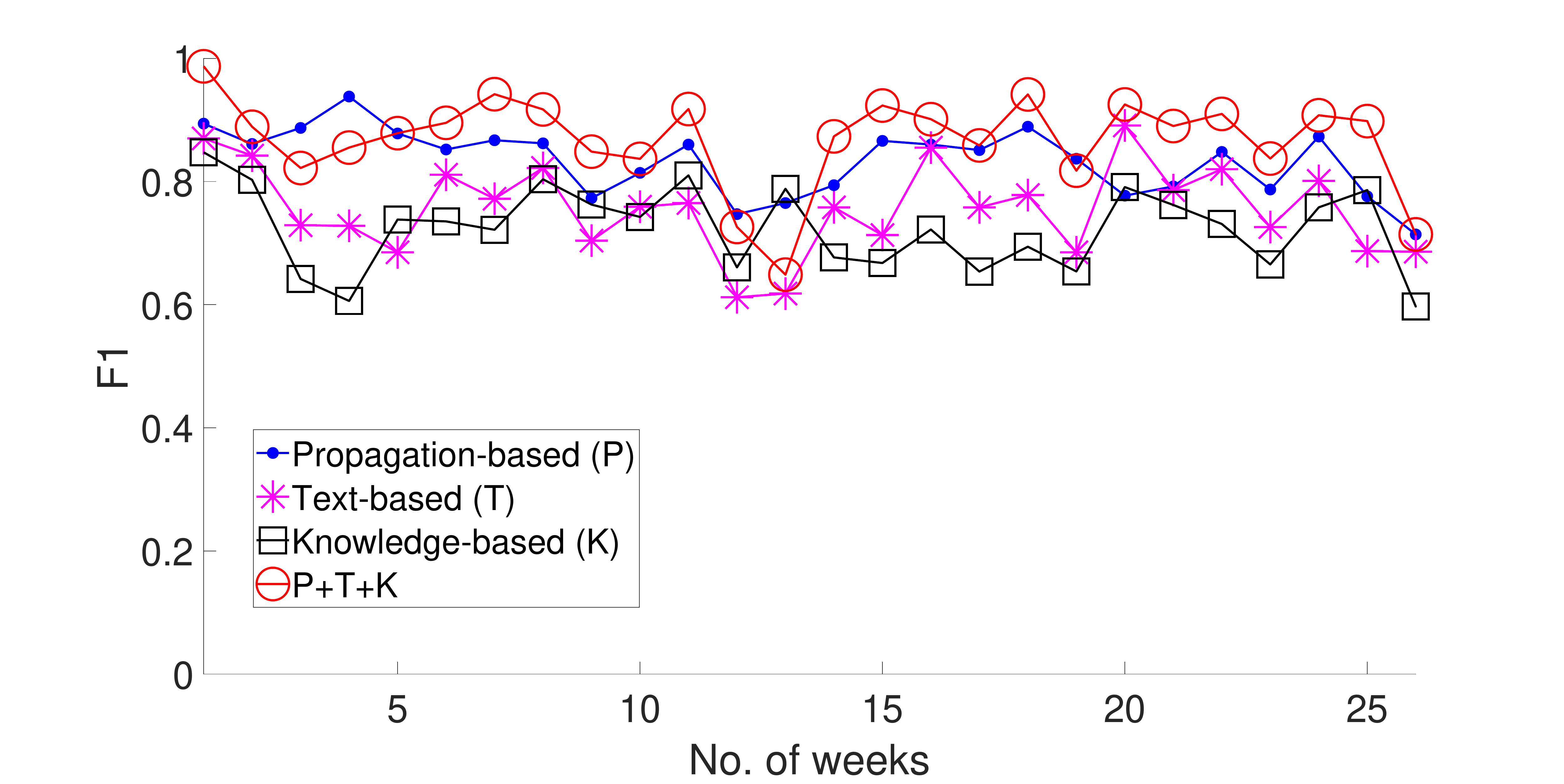}
  \caption{F1}
  \label{figure_f1}
\end{subfigure}
\caption{Time sensitivity analysis for the dataset of GossipCop. The x-axis is the number of weeks from the test item to the last training item.}
\label{figure_time_sensitivity}
\end{figure*}

We run time sensitivity analysis over a period of 26 weeks (half a year) on the dataset of GossipCop  (the dataset of PolitiFact contains only around 400 news items spread over a number of months, which are insufficient to run the analysis): all news items are sorted by their timestamp, and the first 70\% are used for training while the last 30\% are for testing. During test time the data are divided into separate groups based on the number of weeks from the test news item to the last training item, then the four metrics are calculated over each group, \ie the model performance is measured weekly. Fig.~\ref{figure_time_sensitivity} shows how different models perform over time.

We can see that (1) the multi-modal approach performs the best in most cases. (2) In the first few weeks, the performance drop of the text-based and the knowledge-based approaches is more obvious than that of the multi-modal and the propagation-based approaches. (3) All four models are relatively stable between Week 5 and Week 25. The results further confirm the effectiveness of the multi-modal approach that combines extracted knowledge, text content and propagation network.

\section{Related Work}\label{sec:related}
In this section, we provide a brief review of the related work on fake news detection, which has become a popular research problem over recent years. Specifically, following a similar taxonomy in ~\cite{shu_fake_2017,pierri_false_2019}, we classify existing work into three categories: content-based approaches, context-based approaches and mixed approaches, which mainly rely on news content, social context, and a mix of both for detection, respectively. In addition, we also summarise prior work on fake news early detection and explainability.

\subsection{Content-based Approaches}
The most straightforward content-based approach is to consider fake news detection as a text classification problem, and apply techniques such as RST~\cite{rubin_towards_2015}, LIWC~\cite{pennebaker_development_2015} and text-CNN~\cite{kim_convolutional_2014} to identify fake news. These algorithms often serve as baselines.

In addition to the knowledge-based detection methods discussed in this paper, another line of research studies fake news from a style-based perspective. 

\subsubsection{Style-based Detection}
Since the purpose of fake news is to mislead the public, they should exhibit unique writing styles that are rarely seen in real news. This is supported by forensic psychological studies~\cite{undeutsch_beurteilung_1967}, which have shown that statements based on factual experiences differ from those derived from fabrication or fiction in both content and quality.

Therefore, style-based methods aim to identify the different content style, which can be represented by quantifiable features, including attribute-based language features or structure-based language features. For example, Perez-Rosas \etal~\cite{perez-rosas_automatic_2018} train linear SVMs on the following linguistic features to detect fake news: unigrams, bigrams, punctuation, psycholinguistic, readability and syntax features. Other methods that fall into this category include~\cite{horne_this_2017,volkova_separating_2017,wang_liar_2017,potthast_stylometric_2018}.

In addition to textual information, images posted in social media have also been investigated to facilitate the detection of fake news~\cite{jin_novel_2017,yang_ti-cnn_2018,wang_eann_2018,zhou_safe_2020}.

\subsection{Context-based Approaches}
Social context here refers to the interactions between users, including tweet, retweet, reply, mention and follow. As mentioned in the introduction, these engagements can form the propagation pattern for a news item, and a number of studies have used various types of models to identify the difference in the propagation pattern between real and fake news: Wu \etal~\cite{wu_false_2015} use a hybrid SVM; Ma \etal~\cite{ma_detect_2017} use Propagation Tree  Kernel; Wu \etal~\cite{wu_tracing_2018} incorporate LSTM cells into the RNN model; Liu \etal~\cite{liu_early_2018} use both RNNs and CNNs; Shu \etal~\cite{shu_hierarchical_2019} and Zhou \etal~\cite{zhou_network-based_2019} propose different types of features and compare multiple commonly used machine learning models; Monti \etal~\cite{monti_fake_2019}, Lu \etal~\cite{lu_gcan_2020} and Bian \etal~\cite{bian_rumor_2020} apply GNNs to study propagation patterns.

In addition to the propagation network, other types of graphs can also be built from social context. For example, Jin \etal~\cite{jin_news_2016} build a stance network where the weight of an edge represents how much each pair of posts support or deny each other. Then the credibility of all the posts related to a news item is estimated to decide whether the news is fake or real, which can be formalised as a graph optimisation problem. 

In another example, Tacchini \etal~\cite{tacchini_like_2017} propose to detect fake news based on users who liked them on Facebook. They tested logistic regression-based and harmonic Boolean label crowdsourcing-based methods, and their results suggest that both methods can achieve high accuracy.

While all the above methods are supervised, an unsupervised approach is proposed by Yang \etal~\cite{yang_unsupervised_2019}, which builds a Bayesian probability graphical model to capture the generative process among the validity of news, user opinions and user credibility.

\subsection{Mixed Approaches}
Since both news content and social context can provide valuable evidence, mixed approaches use these two sources of information to differentiate between fake news and real news.

Ruchansky \etal~\cite{ruchansky_csi_2017} design a three-module architecture that combines the text of a news article, the received user response and the source of the news: (1) in order to capture temporal representations of articles, the first module trains a RNN that takes the user response, news content and user features as input; (2) the second module generates for each user a score and a low-dimensional representation based on user features; (3) the third module takes the output of the first two modules and trains a neural network to label the news item.

A pre-extracted word set is used in~\cite{zhang_fakedetector_2018} to construct explicit features from the news content, user profile and news subject description. Meanwhile, RNNs are applied to learn latent features, such as news article content information inconsistency and profile latent patterns. Once the features are obtained, a deep diffusive network is built to learn the representations of news articles, creators and subjects.

Shu \etal~\cite{shu_beyond_2019} propose to use the tri-relationship among publishers, news articles and users to detect fake news. Specifically, the latent representations for news content and users are learned with non-negative matrix factorization, and the problem is formalised as an optimisation over the linear combination of each relation. They compare a number of machine learning algorithms to solve the optimisation problem in their experiments.

\subsection{Explainability}
In addition to the above work, a few recent papers have started to work on explainability, which provides evidence to support why their model labels certain news items as fake/real~\cite{popat_declare_2018,shu_defend_2019,lu_gcan_2020}.

For example, Lu \etal~\cite{lu_gcan_2020} propose a novel attention mechanism for fake news detection which jointly considers news textual content, retweet sequences in propagation networks, and user co-occurrence networks.

\subsection{Fake News Early Detection}
Considering that it is difficult to correct people's perception towards an issue, even if the previous impression is inaccurate~\cite{keersmaecker_fake_2017}, it is more crucial to detect fake news at an early stage before it becomes widespread. Therefore, another line of research works on early detection of fake news~\cite{liu_early_2018,shu_leveraging_2020,zhou_fake_2020}. Shu \etal~\cite{shu_leveraging_2020} study multiple weak signals of user sentiment, bias and credibility, and then combine weakly labelled data with a small amount of manually labelled data to train a fake news detection model.

\section{Conclusions and Future Work}\label{sec:conc}
A series of incidents over recent years have demonstrated the significant damage that fake news can cause to society. In this work, we investigate knowledge-enhanced multi-modal techniques for fake news detection. Specifically, we transform the problem of detecting fake news into a subgraph classification task, and design a knowledge-based algorithm that does not require any external knowledge graph. In addition, we propose a multi-modal detection algorithm that combines extracted knowledge, textual content and social context. Experimental results on two datasets with thousands of labelled news items demonstrate the effectiveness of our approaches.

For future work, we will further explore more sophisticated methods rather than simple concatenation to combine different sources of information. Moreover, in addition to news content and propagation networks, we intend to exploit other modalities as well, including images and external knowledge graphs.

\bibliographystyle{ACM-Reference-Format}
\bibliography{references}


\begin{thebibliography}{60}


\ifx \showCODEN    \undefined \def \showCODEN     #1{\unskip}     \fi
\ifx \showDOI      \undefined \def \showDOI       #1{#1}\fi
\ifx \showISBNx    \undefined \def \showISBNx     #1{\unskip}     \fi
\ifx \showISBNxiii \undefined \def \showISBNxiii  #1{\unskip}     \fi
\ifx \showISSN     \undefined \def \showISSN      #1{\unskip}     \fi
\ifx \showLCCN     \undefined \def \showLCCN      #1{\unskip}     \fi
\ifx \shownote     \undefined \def \shownote      #1{#1}          \fi
\ifx \showarticletitle \undefined \def \showarticletitle #1{#1}   \fi
\ifx \showURL      \undefined \def \showURL       {\relax}        \fi
\providecommand\bibfield[2]{#2}
\providecommand\bibinfo[2]{#2}
\providecommand\natexlab[1]{#1}
\providecommand\showeprint[2][]{arXiv:#2}

\bibitem[\protect\citeauthoryear{??}{noa}{1999}]%
        {noauthor_resource_2017}
 \bibinfo{year}{1999}\natexlab{}.
\newblock \bibinfo{booktitle}{\emph{Resource Description Framework ({RDF})
  Model and Syntax Specification}}.
\newblock
\urldef\tempurl%
\url{https://www.w3.org/TR/PR-rdf-syntax/}
\showURL{%
\tempurl}


\bibitem[\protect\citeauthoryear{Alsentzer, Finlayson, Li, and
  Zitnik}{Alsentzer et~al\mbox{.}}{2020}]%
        {alsentzer_subgraph_2020}
\bibfield{author}{\bibinfo{person}{Emily Alsentzer}, \bibinfo{person}{Samuel~G
  Finlayson}, \bibinfo{person}{Michelle~M Li}, {and} \bibinfo{person}{Marinka
  Zitnik}.} \bibinfo{year}{2020}\natexlab{}.
\newblock \showarticletitle{Subgraph Neural Networks}.
\newblock \bibinfo{journal}{\emph{Proceedings of {NeurIPS}}}
  (\bibinfo{year}{2020}).
\newblock


\bibitem[\protect\citeauthoryear{Battaglia, Hamrick, Bapst, Sanchez-Gonzalez,
  Zambaldi, Malinowski, Tacchetti, Raposo, Santoro, Faulkner, Gulcehre, Song,
  Ballard, Gilmer, Dahl, Vaswani, Allen, Nash, Langston, Dyer, Heess, Wierstra,
  Kohli, Botvinick, Vinyals, Li, and Pascanu}{Battaglia et~al\mbox{.}}{2018}]%
        {battaglia_relational_2018}
\bibfield{author}{\bibinfo{person}{Peter~W. Battaglia},
  \bibinfo{person}{Jessica~B. Hamrick}, \bibinfo{person}{Victor Bapst},
  \bibinfo{person}{Alvaro Sanchez-Gonzalez}, \bibinfo{person}{Vinicius
  Zambaldi}, \bibinfo{person}{Mateusz Malinowski}, \bibinfo{person}{Andrea
  Tacchetti}, \bibinfo{person}{David Raposo}, \bibinfo{person}{Adam Santoro},
  \bibinfo{person}{Ryan Faulkner}, \bibinfo{person}{Caglar Gulcehre},
  \bibinfo{person}{Francis Song}, \bibinfo{person}{Andrew Ballard},
  \bibinfo{person}{Justin Gilmer}, \bibinfo{person}{George Dahl},
  \bibinfo{person}{Ashish Vaswani}, \bibinfo{person}{Kelsey Allen},
  \bibinfo{person}{Charles Nash}, \bibinfo{person}{Victoria Langston},
  \bibinfo{person}{Chris Dyer}, \bibinfo{person}{Nicolas Heess},
  \bibinfo{person}{Daan Wierstra}, \bibinfo{person}{Pushmeet Kohli},
  \bibinfo{person}{Matt Botvinick}, \bibinfo{person}{Oriol Vinyals},
  \bibinfo{person}{Yujia Li}, {and} \bibinfo{person}{Razvan Pascanu}.}
  \bibinfo{year}{2018}\natexlab{}.
\newblock \showarticletitle{Relational inductive biases, deep learning, and
  graph networks}.
\newblock \bibinfo{journal}{\emph{eprint {arXiv}}} (\bibinfo{year}{2018}),
  \bibinfo{pages}{arXiv:1806.01261}.
\newblock


\bibitem[\protect\citeauthoryear{Bian, Xiao, Xu, Zhao, Huang, Rong, and
  Huang}{Bian et~al\mbox{.}}{2020}]%
        {bian_rumor_2020}
\bibfield{author}{\bibinfo{person}{Tian Bian}, \bibinfo{person}{Xi Xiao},
  \bibinfo{person}{Tingyang Xu}, \bibinfo{person}{Peilin Zhao},
  \bibinfo{person}{Wenbing Huang}, \bibinfo{person}{Yu Rong}, {and}
  \bibinfo{person}{Junzhou Huang}.} \bibinfo{year}{2020}\natexlab{}.
\newblock \showarticletitle{Rumor Detection on Social Media with Bi-Directional
  Graph Convolutional Networks}.
\newblock  (\bibinfo{year}{2020}), \bibinfo{pages}{arXiv:2001.06362}.
\newblock


\bibitem[\protect\citeauthoryear{Boldi and Rosa}{Boldi and Rosa}{2012}]%
        {boldi_arc-community_2012}
\bibfield{author}{\bibinfo{person}{Paolo Boldi} {and} \bibinfo{person}{Marco
  Rosa}.} \bibinfo{year}{2012}\natexlab{}.
\newblock \showarticletitle{Arc-Community Detection via Triangular Random
  Walks}. In \bibinfo{booktitle}{\emph{2012 Eighth Latin American Web
  Congress}}. \bibinfo{pages}{48--56}.
\newblock


\bibitem[\protect\citeauthoryear{Bordes, Usunier, Garcia-Duran, Weston, and
  Yakhnenko}{Bordes et~al\mbox{.}}{2013}]%
        {bordes_translating_2013}
\bibfield{author}{\bibinfo{person}{Antoine Bordes}, \bibinfo{person}{Nicolas
  Usunier}, \bibinfo{person}{Alberto Garcia-Duran}, \bibinfo{person}{Jason
  Weston}, {and} \bibinfo{person}{Oksana Yakhnenko}.}
  \bibinfo{year}{2013}\natexlab{}.
\newblock \showarticletitle{Translating Embeddings for Modeling
  Multi-relational Data}. In \bibinfo{booktitle}{\emph{Proc. of NeurIPS}}.
\newblock


\bibitem[\protect\citeauthoryear{Cui and Lee}{Cui and Lee}{2020}]%
        {cui2020coaid}
\bibfield{author}{\bibinfo{person}{Limeng Cui} {and} \bibinfo{person}{Dongwon
  Lee}.} \bibinfo{year}{2020}\natexlab{}.
\newblock \showarticletitle{Coaid: Covid-19 healthcare misinformation dataset}.
\newblock \bibinfo{journal}{\emph{arXiv preprint arXiv:2006.00885}}
  (\bibinfo{year}{2020}).
\newblock


\bibitem[\protect\citeauthoryear{Cui, Seo, Tabar, Ma, Wang, and Lee}{Cui
  et~al\mbox{.}}{2020}]%
        {cui_deterrent_2020}
\bibfield{author}{\bibinfo{person}{Limeng Cui}, \bibinfo{person}{Haeseung Seo},
  \bibinfo{person}{Maryam Tabar}, \bibinfo{person}{Fenglong Ma},
  \bibinfo{person}{Suhang Wang}, {and} \bibinfo{person}{Dongwon Lee}.}
  \bibinfo{year}{2020}\natexlab{}.
\newblock \showarticletitle{{DETERRENT}: Knowledge Guided Graph Attention
  Network for Detecting Healthcare Misinformation}. In
  \bibinfo{booktitle}{\emph{Proceedings of the 26th {ACM} {SIGKDD}
  International Conference on Knowledge Discovery \& Data Mining}}
  \emph{(\bibinfo{series}{{KDD} '20})}. \bibinfo{publisher}{Association for
  Computing Machinery}, \bibinfo{pages}{492--502}.
\newblock
\showISBNx{978-1-4503-7998-4}


\bibitem[\protect\citeauthoryear{Hamilton, Ying, and Leskovec}{Hamilton
  et~al\mbox{.}}{2017}]%
        {hamilton_inductive_2017}
\bibfield{author}{\bibinfo{person}{Will Hamilton}, \bibinfo{person}{Zhitao
  Ying}, {and} \bibinfo{person}{Jure Leskovec}.}
  \bibinfo{year}{2017}\natexlab{}.
\newblock \showarticletitle{Inductive Representation Learning on Large Graphs}.
\newblock In \bibinfo{booktitle}{\emph{NeurIPS-2017}}.
  \bibinfo{publisher}{Curran Associates, Inc.}, \bibinfo{pages}{1024--1034}.
\newblock


\bibitem[\protect\citeauthoryear{Hamilton}{Hamilton}{2020}]%
        {hamilton_graph_2020}
\bibfield{author}{\bibinfo{person}{William~L. Hamilton}.}
  \bibinfo{year}{2020}\natexlab{}.
\newblock \showarticletitle{Graph Representation Learning}.
\newblock \bibinfo{journal}{\emph{Synthesis Lectures on Artificial Intelligence
  and Machine Learning}} \bibinfo{volume}{14}, \bibinfo{number}{3}
  (\bibinfo{year}{2020}), \bibinfo{pages}{1--159}.
\newblock


\bibitem[\protect\citeauthoryear{Han, Gao, Yao, Ye, Liu, and Sun}{Han
  et~al\mbox{.}}{2019}]%
        {han_opennre_2019}
\bibfield{author}{\bibinfo{person}{Xu Han}, \bibinfo{person}{Tianyu Gao},
  \bibinfo{person}{Yuan Yao}, \bibinfo{person}{Deming Ye},
  \bibinfo{person}{Zhiyuan Liu}, {and} \bibinfo{person}{Maosong Sun}.}
  \bibinfo{year}{2019}\natexlab{}.
\newblock \showarticletitle{{OpenNRE}: An Open and Extensible Toolkit for
  Neural Relation Extraction}. In \bibinfo{booktitle}{\emph{Proceedings of
  {EMNLP}-{IJCNLP}: System Demonstrations}}. \bibinfo{pages}{169--174}.
\newblock


\bibitem[\protect\citeauthoryear{Han, Karunasekera, and Leckie}{Han
  et~al\mbox{.}}{2020}]%
        {han_graph_2020}
\bibfield{author}{\bibinfo{person}{Yi Han}, \bibinfo{person}{Shanika
  Karunasekera}, {and} \bibinfo{person}{Christopher Leckie}.}
  \bibinfo{year}{2020}\natexlab{}.
\newblock \showarticletitle{Graph Neural Networks with Continual Learning for
  Fake News Detection from Social Media}.
\newblock \bibinfo{journal}{\emph{{arXiv} e-prints}} (\bibinfo{year}{2020}),
  \bibinfo{pages}{arXiv:2007.03316}.
\newblock


\bibitem[\protect\citeauthoryear{Horne and Adali}{Horne and Adali}{2017}]%
        {horne_this_2017}
\bibfield{author}{\bibinfo{person}{Benjamin~D. Horne} {and}
  \bibinfo{person}{Sibel Adali}.} \bibinfo{year}{2017}\natexlab{}.
\newblock \showarticletitle{This Just In: Fake News Packs a Lot in Title, Uses
  Simpler, Repetitive Content in Text Body, More Similar to Satire than Real
  News}.
\newblock \bibinfo{journal}{\emph{{arXiv} e-prints}} (\bibinfo{year}{2017}),
  \bibinfo{pages}{arXiv:1703.09398}.
\newblock


\bibitem[\protect\citeauthoryear{Jin, Cao, Zhang, and Luo}{Jin
  et~al\mbox{.}}{2016}]%
        {jin_news_2016}
\bibfield{author}{\bibinfo{person}{Zhiwei Jin}, \bibinfo{person}{Juan Cao},
  \bibinfo{person}{Yongdong Zhang}, {and} \bibinfo{person}{Jiebo Luo}.}
  \bibinfo{year}{2016}\natexlab{}.
\newblock \showarticletitle{News Verification by Exploiting Conflicting Social
  Viewpoints in Microblogs}. In \bibinfo{booktitle}{\emph{Proceedings of the
  30th {AAAI}}} \emph{(\bibinfo{series}{{AAAI}’16})}.
  \bibinfo{address}{Phoenix, Arizona}, \bibinfo{pages}{2972--2978}.
\newblock


\bibitem[\protect\citeauthoryear{Jin, Cao, Zhang, Zhou, and Tian}{Jin
  et~al\mbox{.}}{2017}]%
        {jin_novel_2017}
\bibfield{author}{\bibinfo{person}{Z. Jin}, \bibinfo{person}{J. Cao},
  \bibinfo{person}{Y. Zhang}, \bibinfo{person}{J. Zhou}, {and}
  \bibinfo{person}{Q. Tian}.} \bibinfo{year}{2017}\natexlab{}.
\newblock \showarticletitle{Novel Visual and Statistical Image Features for
  Microblogs News Verification}.
\newblock \bibinfo{journal}{\emph{{IEEE} Transactions on Multimedia}}
  \bibinfo{volume}{19}, \bibinfo{number}{3} (\bibinfo{year}{2017}),
  \bibinfo{pages}{598--608}.
\newblock


\bibitem[\protect\citeauthoryear{keersmaecker and Roets}{keersmaecker and
  Roets}{2017}]%
        {keersmaecker_fake_2017}
\bibfield{author}{\bibinfo{person}{Jonas~De keersmaecker} {and}
  \bibinfo{person}{Arne Roets}.} \bibinfo{year}{2017}\natexlab{}.
\newblock \showarticletitle{‘Fake news’: Incorrect, but hard to correct.
  The role of cognitive ability on the impact of false information on social
  impressions}.
\newblock \bibinfo{journal}{\emph{Intelligence}}  \bibinfo{volume}{65}
  (\bibinfo{year}{2017}), \bibinfo{pages}{107 -- 110}.
\newblock
\showISSN{0160-2896}


\bibitem[\protect\citeauthoryear{Kim}{Kim}{2014}]%
        {kim_convolutional_2014}
\bibfield{author}{\bibinfo{person}{Yoon Kim}.} \bibinfo{year}{2014}\natexlab{}.
\newblock \showarticletitle{Convolutional Neural Networks for Sentence
  Classification}. In \bibinfo{booktitle}{\emph{Proceedings of the 2014
  Conference on Empirical Methods in Natural Language Processing ({EMNLP})}}
  (Doha, Qatar). \bibinfo{pages}{1746--1751}.
\newblock


\bibitem[\protect\citeauthoryear{Kochkina, Liakata, and Augenstein}{Kochkina
  et~al\mbox{.}}{2017}]%
        {kochkina2017turing}
\bibfield{author}{\bibinfo{person}{Elena Kochkina}, \bibinfo{person}{Maria
  Liakata}, {and} \bibinfo{person}{Isabelle Augenstein}.}
  \bibinfo{year}{2017}\natexlab{}.
\newblock \showarticletitle{{Turing at SemEval-2017 Task 8: Sequential Approach
  to Rumour Stance Classification with Branch-LSTM}}. In
  \bibinfo{booktitle}{\emph{Proc. of SemEval}}.
\newblock


\bibitem[\protect\citeauthoryear{Liu and Wu}{Liu and Wu}{2018}]%
        {liu_early_2018}
\bibfield{author}{\bibinfo{person}{Yang Liu} {and}
  \bibinfo{person}{Yi-fang~Brook Wu}.} \bibinfo{year}{2018}\natexlab{}.
\newblock \showarticletitle{Early Detection of Fake News on Social Media
  Through Propagation Path Classification with Recurrent and Convolutional
  Networks}. In \bibinfo{booktitle}{\emph{Proceedings of the 32nd {AAAI}}}.
  \bibinfo{pages}{354--361}.
\newblock


\bibitem[\protect\citeauthoryear{Lu and Li}{Lu and Li}{2020}]%
        {lu_gcan_2020}
\bibfield{author}{\bibinfo{person}{Yi-Ju Lu} {and} \bibinfo{person}{Cheng-Te
  Li}.} \bibinfo{year}{2020}\natexlab{}.
\newblock \showarticletitle{{GCAN}: Graph-aware Co-Attention Networks for
  Explainable Fake News Detection on Social Media}.
\newblock  (\bibinfo{year}{2020}), \bibinfo{pages}{arXiv:2004.11648}.
\newblock


\bibitem[\protect\citeauthoryear{Ma, Gao, Mitra, Kwon, Jansen, Wong, and
  Cha}{Ma et~al\mbox{.}}{2016}]%
        {ma2016detecting}
\bibfield{author}{\bibinfo{person}{Jing Ma}, \bibinfo{person}{Wei Gao},
  \bibinfo{person}{Prasenjit Mitra}, \bibinfo{person}{Sejeong Kwon},
  \bibinfo{person}{Bernard~J Jansen}, \bibinfo{person}{Kam-Fai Wong}, {and}
  \bibinfo{person}{Meeyoung Cha}.} \bibinfo{year}{2016}\natexlab{}.
\newblock \showarticletitle{Detecting rumors from microblogs with recurrent
  neural networks}. In \bibinfo{booktitle}{\emph{Proc. of IJCAI}}.
\newblock


\bibitem[\protect\citeauthoryear{Ma, Gao, and Wong}{Ma et~al\mbox{.}}{2017}]%
        {ma_detect_2017}
\bibfield{author}{\bibinfo{person}{Jing Ma}, \bibinfo{person}{Wei Gao}, {and}
  \bibinfo{person}{Kam-Fai Wong}.} \bibinfo{year}{2017}\natexlab{}.
\newblock \showarticletitle{Detect Rumors in Microblog Posts Using Propagation
  Structure via Kernel Learning}. In \bibinfo{booktitle}{\emph{Proceedings of
  the 55th ACL}} (Vancouver, Canada). \bibinfo{pages}{708--717}.
\newblock


\bibitem[\protect\citeauthoryear{Monti, Frasca, Eynard, Mannion, and
  Bronstein}{Monti et~al\mbox{.}}{2019}]%
        {monti_fake_2019}
\bibfield{author}{\bibinfo{person}{Federico Monti}, \bibinfo{person}{Fabrizio
  Frasca}, \bibinfo{person}{Davide Eynard}, \bibinfo{person}{Damon Mannion},
  {and} \bibinfo{person}{Michael~M. Bronstein}.}
  \bibinfo{year}{2019}\natexlab{}.
\newblock \showarticletitle{Fake News Detection on Social Media using Geometric
  Deep Learning}.
\newblock \bibinfo{journal}{\emph{{arXiv} e-prints}} (\bibinfo{year}{2019}),
  \bibinfo{pages}{arXiv:1902.06673}.
\newblock


\bibitem[\protect\citeauthoryear{Pan, Pavlova, Li, Li, Li, and Liu}{Pan
  et~al\mbox{.}}{2018}]%
        {pan_content_2018}
\bibfield{author}{\bibinfo{person}{Jeff~Z. Pan}, \bibinfo{person}{Siyana
  Pavlova}, \bibinfo{person}{Chenxi Li}, \bibinfo{person}{Ningxi Li},
  \bibinfo{person}{Yangmei Li}, {and} \bibinfo{person}{Jinshuo Liu}.}
  \bibinfo{year}{2018}\natexlab{}.
\newblock \showarticletitle{Content Based Fake News Detection Using Knowledge
  Graphs}. In \bibinfo{booktitle}{\emph{The Semantic Web – {ISWC} 2018}}
  (Cham). \bibinfo{publisher}{Springer International Publishing},
  \bibinfo{pages}{669--683}.
\newblock
\showISBNx{978-3-030-00671-6}


\bibitem[\protect\citeauthoryear{Pennebaker, Boyd, Jordan, and
  Blackburn}{Pennebaker et~al\mbox{.}}{2015}]%
        {pennebaker_development_2015}
\bibfield{author}{\bibinfo{person}{James~W. Pennebaker},
  \bibinfo{person}{Ryan~L. Boyd}, \bibinfo{person}{Kayla Jordan}, {and}
  \bibinfo{person}{Kate Blackburn}.} \bibinfo{year}{2015}\natexlab{}.
\newblock \bibinfo{booktitle}{\emph{The Development and Psychometric Properties
  of {LIWC} 2015}}.
\newblock \bibinfo{type}{{T}echnical {R}eport}.
\newblock
\urldef\tempurl%
\url{https://repositories.lib.utexas.edu/handle/2152/31333}
\showURL{%
\tempurl}


\bibitem[\protect\citeauthoryear{Pierri and Ceri}{Pierri and Ceri}{2019}]%
        {pierri_false_2019}
\bibfield{author}{\bibinfo{person}{Francesco Pierri} {and}
  \bibinfo{person}{Stefano Ceri}.} \bibinfo{year}{2019}\natexlab{}.
\newblock \showarticletitle{False News On Social Media: A Data-Driven Survey}.
\newblock \bibinfo{journal}{\emph{{SIGMOD} Record}} \bibinfo{volume}{48},
  \bibinfo{number}{2} (\bibinfo{year}{2019}), \bibinfo{pages}{18--27}.
\newblock
\showISSN{0163-5808}


\bibitem[\protect\citeauthoryear{Popat, Mukherjee, Yates, and Weikum}{Popat
  et~al\mbox{.}}{2018}]%
        {popat_declare_2018}
\bibfield{author}{\bibinfo{person}{Kashyap Popat}, \bibinfo{person}{Subhabrata
  Mukherjee}, \bibinfo{person}{Andrew Yates}, {and} \bibinfo{person}{Gerhard
  Weikum}.} \bibinfo{year}{2018}\natexlab{}.
\newblock \showarticletitle{{DeClarE}: Debunking Fake News and False Claims
  using Evidence-Aware Deep Learning}. In \bibinfo{booktitle}{\emph{Proceedings
  of the 2018 Conference on Empirical Methods in Natural Language Processing}}
  (Brussels, Belgium). \bibinfo{pages}{22--32}.
\newblock


\bibitem[\protect\citeauthoryear{Potthast, Kiesel, Reinartz, Bevendorff, and
  Stein}{Potthast et~al\mbox{.}}{2018}]%
        {potthast_stylometric_2018}
\bibfield{author}{\bibinfo{person}{Martin Potthast}, \bibinfo{person}{Johannes
  Kiesel}, \bibinfo{person}{Kevin Reinartz}, \bibinfo{person}{Janek
  Bevendorff}, {and} \bibinfo{person}{Benno Stein}.}
  \bibinfo{year}{2018}\natexlab{}.
\newblock \showarticletitle{A Stylometric Inquiry into Hyperpartisan and Fake
  News}. In \bibinfo{booktitle}{\emph{Proceedings of the 56th ACL}} (Melbourne,
  Australia). \bibinfo{pages}{231--240}.
\newblock


\bibitem[\protect\citeauthoryear{Pérez-Rosas, Kleinberg, Lefevre, and
  Mihalcea}{Pérez-Rosas et~al\mbox{.}}{2018}]%
        {perez-rosas_automatic_2018}
\bibfield{author}{\bibinfo{person}{Verónica Pérez-Rosas},
  \bibinfo{person}{Bennett Kleinberg}, \bibinfo{person}{Alexandra Lefevre},
  {and} \bibinfo{person}{Rada Mihalcea}.} \bibinfo{year}{2018}\natexlab{}.
\newblock \showarticletitle{Automatic Detection of Fake News}. In
  \bibinfo{booktitle}{\emph{Proceedings of the 27th International Conference on
  Computational Linguistics}}. \bibinfo{pages}{3391--3401}.
\newblock


\bibitem[\protect\citeauthoryear{Rubin, Conroy, and Chen}{Rubin
  et~al\mbox{.}}{2015}]%
        {rubin_towards_2015}
\bibfield{author}{\bibinfo{person}{Victoria Rubin}, \bibinfo{person}{Nadia
  Conroy}, {and} \bibinfo{person}{Yimin Chen}.}
  \bibinfo{year}{2015}\natexlab{}.
\newblock \showarticletitle{Towards News Verification: Deception Detection
  Methods for News Discourse}. In \bibinfo{booktitle}{\emph{Proceedings of the
  48th Hawaii International Conference on System Sciences ({HICSS}48)}}.
\newblock


\bibitem[\protect\citeauthoryear{Ruchansky, Seo, and Liu}{Ruchansky
  et~al\mbox{.}}{2017}]%
        {ruchansky_csi_2017}
\bibfield{author}{\bibinfo{person}{Natali Ruchansky}, \bibinfo{person}{Sungyong
  Seo}, {and} \bibinfo{person}{Yan Liu}.} \bibinfo{year}{2017}\natexlab{}.
\newblock \showarticletitle{{CSI}: A Hybrid Deep Model for Fake News
  Detection}. In \bibinfo{booktitle}{\emph{Proceedings of the 2017 {ACM} on
  Conference on Information and Knowledge Management}}
  \emph{(\bibinfo{series}{{CIKM} ’17})}. \bibinfo{pages}{797--806}.
\newblock
\showISBNx{978-1-4503-4918-5}


\bibitem[\protect\citeauthoryear{Santia and Williams}{Santia and
  Williams}{2018}]%
        {santia2018buzzface}
\bibfield{author}{\bibinfo{person}{Giovanni Santia} {and} \bibinfo{person}{Jake
  Williams}.} \bibinfo{year}{2018}\natexlab{}.
\newblock \showarticletitle{Buzzface: A news veracity dataset with facebook
  user commentary and egos}. In \bibinfo{booktitle}{\emph{Proc. of ICWSM}}.
\newblock


\bibitem[\protect\citeauthoryear{Shu, Cui, Wang, Lee, and Liu}{Shu
  et~al\mbox{.}}{2019a}]%
        {shu_defend_2019}
\bibfield{author}{\bibinfo{person}{Kai Shu}, \bibinfo{person}{Limeng Cui},
  \bibinfo{person}{Suhang Wang}, \bibinfo{person}{Dongwon Lee}, {and}
  \bibinfo{person}{Huan Liu}.} \bibinfo{year}{2019}\natexlab{a}.
\newblock \showarticletitle{{DEFEND}: Explainable Fake News Detection}. In
  \bibinfo{booktitle}{\emph{Proceedings of the 25th {ACM} {SIGKDD}
  International Conference on Knowledge Discovery \& Data Mining}} (New York,
  {NY}, {USA}) \emph{(\bibinfo{series}{{KDD} ’19})}.
  \bibinfo{address}{Anchorage, {AK}, {USA}}, \bibinfo{pages}{395--405}.
\newblock
\showISBNx{978-1-4503-6201-6}


\bibitem[\protect\citeauthoryear{Shu, Mahudeswaran, Wang, Lee, and Liu}{Shu
  et~al\mbox{.}}{2018}]%
        {shu_fakenewsnet_2018}
\bibfield{author}{\bibinfo{person}{Kai Shu}, \bibinfo{person}{Deepak
  Mahudeswaran}, \bibinfo{person}{Suhang Wang}, \bibinfo{person}{Dongwon Lee},
  {and} \bibinfo{person}{Huan Liu}.} \bibinfo{year}{2018}\natexlab{}.
\newblock \showarticletitle{{FakeNewsNet}: A Data Repository with News Content,
  Social Context and Spatialtemporal Information for Studying Fake News on
  Social Media}.
\newblock \bibinfo{journal}{\emph{{arXiv} e-prints}} (\bibinfo{year}{2018}),
  \bibinfo{pages}{arXiv:1809.01286}.
\newblock


\bibitem[\protect\citeauthoryear{Shu, Mahudeswaran, Wang, and Liu}{Shu
  et~al\mbox{.}}{2019b}]%
        {shu_hierarchical_2019}
\bibfield{author}{\bibinfo{person}{Kai Shu}, \bibinfo{person}{Deepak
  Mahudeswaran}, \bibinfo{person}{Suhang Wang}, {and} \bibinfo{person}{Huan
  Liu}.} \bibinfo{year}{2019}\natexlab{b}.
\newblock \showarticletitle{Hierarchical Propagation Networks for Fake News
  Detection: Investigation and Exploitation}.
\newblock \bibinfo{journal}{\emph{{arXiv} e-prints}} (\bibinfo{year}{2019}),
  \bibinfo{pages}{arXiv:1903.09196}.
\newblock


\bibitem[\protect\citeauthoryear{Shu, Sliva, Wang, Tang, and Liu}{Shu
  et~al\mbox{.}}{2017}]%
        {shu_fake_2017}
\bibfield{author}{\bibinfo{person}{Kai Shu}, \bibinfo{person}{Amy Sliva},
  \bibinfo{person}{Suhang Wang}, \bibinfo{person}{Jiliang Tang}, {and}
  \bibinfo{person}{Huan Liu}.} \bibinfo{year}{2017}\natexlab{}.
\newblock \showarticletitle{Fake News Detection on Social Media: A Data Mining
  Perspective}.
\newblock \bibinfo{journal}{\emph{{SIGKDD} Explorations Newsletter}}
  \bibinfo{volume}{19}, \bibinfo{number}{1} (\bibinfo{year}{2017}),
  \bibinfo{pages}{22--36}.
\newblock
\showISSN{1931-0145}


\bibitem[\protect\citeauthoryear{Shu, Wang, and Liu}{Shu
  et~al\mbox{.}}{2019c}]%
        {shu_beyond_2019}
\bibfield{author}{\bibinfo{person}{Kai Shu}, \bibinfo{person}{Suhang Wang},
  {and} \bibinfo{person}{Huan Liu}.} \bibinfo{year}{2019}\natexlab{c}.
\newblock \showarticletitle{Beyond News Contents: The Role of Social Context
  for Fake News Detection}. In \bibinfo{booktitle}{\emph{Proceedings of the
  Twelfth {ACM} International Conference on Web Search and Data Mining}} (New
  York, {NY}, {USA}) \emph{(\bibinfo{series}{{WSDM} ’19})}.
  \bibinfo{address}{Melbourne, {VIC}, Australia}, \bibinfo{pages}{312--320}.
\newblock
\showISBNx{978-1-4503-5940-5}


\bibitem[\protect\citeauthoryear{Shu, Zheng, Li, Mukherjee, Hassan~Awadallah,
  Ruston, and Liu}{Shu et~al\mbox{.}}{2020}]%
        {shu_leveraging_2020}
\bibfield{author}{\bibinfo{person}{Kai Shu}, \bibinfo{person}{Guoqing Zheng},
  \bibinfo{person}{Yichuan Li}, \bibinfo{person}{Subhabrata Mukherjee},
  \bibinfo{person}{Ahmed Hassan~Awadallah}, \bibinfo{person}{Scott Ruston},
  {and} \bibinfo{person}{Huan Liu}.} \bibinfo{year}{2020}\natexlab{}.
\newblock \showarticletitle{Leveraging Multi-Source Weak Social Supervision for
  Early Detection of Fake News}.
\newblock \bibinfo{journal}{\emph{{arXiv}:2004.01732}} (\bibinfo{year}{2020}).
\newblock


\bibitem[\protect\citeauthoryear{Tacchini, Ballarin, Della~Vedova, Moret, and
  de~Alfaro}{Tacchini et~al\mbox{.}}{2017}]%
        {tacchini_like_2017}
\bibfield{author}{\bibinfo{person}{Eugenio Tacchini}, \bibinfo{person}{Gabriele
  Ballarin}, \bibinfo{person}{Marco~L. Della~Vedova}, \bibinfo{person}{Stefano
  Moret}, {and} \bibinfo{person}{Luca de Alfaro}.}
  \bibinfo{year}{2017}\natexlab{}.
\newblock \showarticletitle{Some Like it Hoax: Automated Fake News Detection in
  Social Networks}.
\newblock \bibinfo{journal}{\emph{{arXiv} e-prints}} (\bibinfo{year}{2017}),
  \bibinfo{pages}{arXiv:1704.07506}.
\newblock


\bibitem[\protect\citeauthoryear{Undeutsch}{Undeutsch}{1967}]%
        {undeutsch_beurteilung_1967}
\bibfield{author}{\bibinfo{person}{Udo Undeutsch}.}
  \bibinfo{year}{1967}\natexlab{}.
\newblock \showarticletitle{Beurteilung der Glaubhaftigkeit von Aussagen}.
\newblock \bibinfo{journal}{\emph{Handbuch der Psychologie, Band 11:
  Forensische Psychologie}} (\bibinfo{year}{1967}), \bibinfo{pages}{26--181}.
\newblock


\bibitem[\protect\citeauthoryear{Volkova, Shaffer, Jang, and Hodas}{Volkova
  et~al\mbox{.}}{2017}]%
        {volkova_separating_2017}
\bibfield{author}{\bibinfo{person}{Svitlana Volkova}, \bibinfo{person}{Kyle
  Shaffer}, \bibinfo{person}{Jin~Yea Jang}, {and} \bibinfo{person}{Nathan
  Hodas}.} \bibinfo{year}{2017}\natexlab{}.
\newblock \showarticletitle{Separating Facts from Fiction: Linguistic Models to
  Classify Suspicious and Trusted News Posts on Twitter}. In
  \bibinfo{booktitle}{\emph{Proceedings of the 55th ACL}} (Vancouver, Canada).
  \bibinfo{pages}{647--653}.
\newblock


\bibitem[\protect\citeauthoryear{Vosoughi, Roy, and Aral}{Vosoughi
  et~al\mbox{.}}{2018}]%
        {vosoughi_spread_2018}
\bibfield{author}{\bibinfo{person}{Soroush Vosoughi}, \bibinfo{person}{Deb
  Roy}, {and} \bibinfo{person}{Sinan Aral}.} \bibinfo{year}{2018}\natexlab{}.
\newblock \showarticletitle{The spread of true and false news online}.
\newblock \bibinfo{journal}{\emph{Science}} \bibinfo{volume}{359},
  \bibinfo{number}{6380} (\bibinfo{year}{2018}), \bibinfo{pages}{1146--1151}.
\newblock
\showISSN{0036-8075}


\bibitem[\protect\citeauthoryear{Wang}{Wang}{2017}]%
        {wang_liar_2017}
\bibfield{author}{\bibinfo{person}{William~Yang Wang}.}
  \bibinfo{year}{2017}\natexlab{}.
\newblock \showarticletitle{“Liar, Liar Pants on Fire”: A New Benchmark
  Dataset for Fake News Detection}. In \bibinfo{booktitle}{\emph{Proceedings of
  the 55th ACL}}. \bibinfo{pages}{422--426}.
\newblock


\bibitem[\protect\citeauthoryear{Wang, Ma, Jin, Yuan, Xun, Jha, Su, and
  Gao}{Wang et~al\mbox{.}}{2018}]%
        {wang_eann_2018}
\bibfield{author}{\bibinfo{person}{Yaqing Wang}, \bibinfo{person}{Fenglong Ma},
  \bibinfo{person}{Zhiwei Jin}, \bibinfo{person}{Ye Yuan},
  \bibinfo{person}{Guangxu Xun}, \bibinfo{person}{Kishlay Jha},
  \bibinfo{person}{Lu Su}, {and} \bibinfo{person}{Jing Gao}.}
  \bibinfo{year}{2018}\natexlab{}.
\newblock \showarticletitle{{EANN}: Event Adversarial Neural Networks for
  Multi-Modal Fake News Detection}. In \bibinfo{booktitle}{\emph{Proceedings of
  the 24th {ACM} {SIGKDD} International Conference on Knowledge Discovery \&
  Data Mining}} (New York, {NY}, {USA}) \emph{(\bibinfo{series}{{KDD} ’18})}.
  \bibinfo{address}{London, United Kingdom}, \bibinfo{pages}{849--857}.
\newblock
\showISBNx{978-1-4503-5552-0}


\bibitem[\protect\citeauthoryear{Wu, Yang, and Zhu}{Wu et~al\mbox{.}}{2015}]%
        {wu_false_2015}
\bibfield{author}{\bibinfo{person}{K. Wu}, \bibinfo{person}{S. Yang}, {and}
  \bibinfo{person}{K.~Q. Zhu}.} \bibinfo{year}{2015}\natexlab{}.
\newblock \showarticletitle{False rumors detection on Sina Weibo by propagation
  structures}. In \bibinfo{booktitle}{\emph{2015 {IEEE} 31st International
  Conference on Data Engineering}}. \bibinfo{pages}{651--662}.
\newblock
\urldef\tempurl%
\url{https://doi.org/10.1109/ICDE.2015.7113322}
\showDOI{\tempurl}


\bibitem[\protect\citeauthoryear{Wu and Liu}{Wu and Liu}{2018}]%
        {wu_tracing_2018}
\bibfield{author}{\bibinfo{person}{Liang Wu} {and} \bibinfo{person}{Huan Liu}.}
  \bibinfo{year}{2018}\natexlab{}.
\newblock \showarticletitle{Tracing Fake-News Footprints: Characterizing Social
  Media Messages by How They Propagate}. In
  \bibinfo{booktitle}{\emph{Proceedings of the Eleventh {ACM} International
  Conference on Web Search and Data Mining}} (New York, {NY}, {USA})
  \emph{(\bibinfo{series}{{WSDM} ’18})}. \bibinfo{address}{Marina Del Rey,
  {CA}, {USA}}, \bibinfo{pages}{637--645}.
\newblock
\showISBNx{978-1-4503-5581-0}


\bibitem[\protect\citeauthoryear{Wu, Pan, Chen, Long, Zhang, and Yu}{Wu
  et~al\mbox{.}}{2019}]%
        {wu_comprehensive_2019}
\bibfield{author}{\bibinfo{person}{Zonghan Wu}, \bibinfo{person}{Shirui Pan},
  \bibinfo{person}{Fengwen Chen}, \bibinfo{person}{Guodong Long},
  \bibinfo{person}{Chengqi Zhang}, {and} \bibinfo{person}{Philip~S. Yu}.}
  \bibinfo{year}{2019}\natexlab{}.
\newblock \showarticletitle{A Comprehensive Survey on Graph Neural Networks}.
\newblock \bibinfo{journal}{\emph{{arXiv} e-prints}} (\bibinfo{year}{2019}),
  \bibinfo{pages}{arXiv:1901.00596}.
\newblock


\bibitem[\protect\citeauthoryear{Yang, Shu, Wang, Gu, Wu, and Liu}{Yang
  et~al\mbox{.}}{2019}]%
        {yang_unsupervised_2019}
\bibfield{author}{\bibinfo{person}{Shuo Yang}, \bibinfo{person}{Kai Shu},
  \bibinfo{person}{Suhang Wang}, \bibinfo{person}{Renjie Gu},
  \bibinfo{person}{Fan Wu}, {and} \bibinfo{person}{Huan Liu}.}
  \bibinfo{year}{2019}\natexlab{}.
\newblock \showarticletitle{Unsupervised Fake News Detection on Social Media: A
  Generative Approach}.
\newblock \bibinfo{journal}{\emph{Proceedings of the 33rd {AAAI}}}
  \bibinfo{volume}{33} (\bibinfo{year}{2019}), \bibinfo{pages}{5644--5651}.
\newblock


\bibitem[\protect\citeauthoryear{Yang, Zheng, Zhang, Cui, Li, and Yu}{Yang
  et~al\mbox{.}}{2018}]%
        {yang_ti-cnn_2018}
\bibfield{author}{\bibinfo{person}{Yang Yang}, \bibinfo{person}{Lei Zheng},
  \bibinfo{person}{Jiawei Zhang}, \bibinfo{person}{Qingcai Cui},
  \bibinfo{person}{Zhoujun Li}, {and} \bibinfo{person}{Philip~S. Yu}.}
  \bibinfo{year}{2018}\natexlab{}.
\newblock \showarticletitle{{TI}-{CNN}: Convolutional Neural Networks for Fake
  News Detection}.
\newblock \bibinfo{journal}{\emph{{arXiv} e-prints}} (\bibinfo{year}{2018}),
  \bibinfo{pages}{arXiv:1806.00749}.
\newblock


\bibitem[\protect\citeauthoryear{Yang, Yang, Dyer, He, Smola, and Hovy}{Yang
  et~al\mbox{.}}{2016}]%
        {yang_hierarchical_2016}
\bibfield{author}{\bibinfo{person}{Zichao Yang}, \bibinfo{person}{Diyi Yang},
  \bibinfo{person}{Chris Dyer}, \bibinfo{person}{Xiaodong He},
  \bibinfo{person}{Alex Smola}, {and} \bibinfo{person}{Eduard Hovy}.}
  \bibinfo{year}{2016}\natexlab{}.
\newblock \showarticletitle{Hierarchical Attention Networks for Document
  Classification}. In \bibinfo{booktitle}{\emph{Proceedings of the 2016
  Conference of the North American Chapter of the Association for Computational
  Linguistics: Human Language Technologies}}. \bibinfo{pages}{1480--1489}.
\newblock


\bibitem[\protect\citeauthoryear{Yao, Ye, Li, Han, Lin, Liu, Liu, Huang, Zhou,
  and Sun}{Yao et~al\mbox{.}}{2019}]%
        {yao_docred_2019}
\bibfield{author}{\bibinfo{person}{Yuan Yao}, \bibinfo{person}{Deming Ye},
  \bibinfo{person}{Peng Li}, \bibinfo{person}{Xu Han}, \bibinfo{person}{Yankai
  Lin}, \bibinfo{person}{Zhenghao Liu}, \bibinfo{person}{Zhiyuan Liu},
  \bibinfo{person}{Lixin Huang}, \bibinfo{person}{Jie Zhou}, {and}
  \bibinfo{person}{Maosong Sun}.} \bibinfo{year}{2019}\natexlab{}.
\newblock \showarticletitle{{DocRED}: A Large-Scale Document-Level Relation
  Extraction Dataset}. In \bibinfo{booktitle}{\emph{Proceedings of {ACL}
  2019}}.
\newblock


\bibitem[\protect\citeauthoryear{Ying, You, Morris, Ren, Hamilton, and
  Leskovec}{Ying et~al\mbox{.}}{2018}]%
        {ying_hierarchical_2018}
\bibfield{author}{\bibinfo{person}{Rex Ying}, \bibinfo{person}{Jiaxuan You},
  \bibinfo{person}{Christopher Morris}, \bibinfo{person}{Xiang Ren},
  \bibinfo{person}{William~L. Hamilton}, {and} \bibinfo{person}{Jure
  Leskovec}.} \bibinfo{year}{2018}\natexlab{}.
\newblock \showarticletitle{Hierarchical Graph Representation Learning with
  Differentiable Pooling}. In \bibinfo{booktitle}{\emph{Proceedings of the 32nd
  NeurIPS}}. \bibinfo{pages}{4805--4815}.
\newblock


\bibitem[\protect\citeauthoryear{Zhang, Dong, and Yu}{Zhang
  et~al\mbox{.}}{2018}]%
        {zhang_fakedetector_2018}
\bibfield{author}{\bibinfo{person}{Jiawei Zhang}, \bibinfo{person}{Bowen Dong},
  {and} \bibinfo{person}{Philip~S. Yu}.} \bibinfo{year}{2018}\natexlab{}.
\newblock \showarticletitle{{FAKEDETECTOR}: Effective Fake News Detection with
  Deep Diffusive Neural Network}.
\newblock \bibinfo{journal}{\emph{{arXiv} e-prints}} (\bibinfo{year}{2018}),
  \bibinfo{pages}{arXiv:1805.08751}.
\newblock


\bibitem[\protect\citeauthoryear{Zhang, Zhong, Chen, Angeli, and Manning}{Zhang
  et~al\mbox{.}}{2017}]%
        {zhang_position-aware_2017}
\bibfield{author}{\bibinfo{person}{Yuhao Zhang}, \bibinfo{person}{Victor
  Zhong}, \bibinfo{person}{Danqi Chen}, \bibinfo{person}{Gabor Angeli}, {and}
  \bibinfo{person}{Christopher~D. Manning}.} \bibinfo{year}{2017}\natexlab{}.
\newblock \showarticletitle{Position-aware Attention and Supervised Data
  Improve Slot Filling}. In \bibinfo{booktitle}{\emph{Proceedings of the 2017
  Conference on Empirical Methods in Natural Language Processing}}.
  \bibinfo{publisher}{Association for Computational Linguistics},
  \bibinfo{pages}{35--45}.
\newblock


\bibitem[\protect\citeauthoryear{Zhou, Huang, Ma, and Huang}{Zhou
  et~al\mbox{.}}{2021}]%
        {zhou_document-level_2021}
\bibfield{author}{\bibinfo{person}{Wenxuan Zhou}, \bibinfo{person}{Kevin
  Huang}, \bibinfo{person}{Tengyu Ma}, {and} \bibinfo{person}{Jing Huang}.}
  \bibinfo{year}{2021}\natexlab{}.
\newblock \showarticletitle{Document-Level Relation Extraction with Adaptive
  Thresholding and Localized Context Pooling}. In
  \bibinfo{booktitle}{\emph{Proceedings of the {AAAI} Conference on Artificial
  Intelligence}}.
\newblock


\bibitem[\protect\citeauthoryear{Zhou, Jain, Phoha, and Zafarani}{Zhou
  et~al\mbox{.}}{2020a}]%
        {zhou_fake_2020}
\bibfield{author}{\bibinfo{person}{Xinyi Zhou}, \bibinfo{person}{Atishay Jain},
  \bibinfo{person}{Vir~V. Phoha}, {and} \bibinfo{person}{Reza Zafarani}.}
  \bibinfo{year}{2020}\natexlab{a}.
\newblock \showarticletitle{Fake News Early Detection: A Theory-Driven Model}.
\newblock \bibinfo{journal}{\emph{Digital Threats: Research and Practice}}
  \bibinfo{volume}{1}, \bibinfo{number}{2} (\bibinfo{year}{2020}),
  \bibinfo{pages}{1--25}.
\newblock


\bibitem[\protect\citeauthoryear{Zhou, Mulay, Ferrara, and Zafarani}{Zhou
  et~al\mbox{.}}{2020b}]%
        {zhou2020recovery}
\bibfield{author}{\bibinfo{person}{Xinyi Zhou}, \bibinfo{person}{Apurva Mulay},
  \bibinfo{person}{Emilio Ferrara}, {and} \bibinfo{person}{Reza Zafarani}.}
  \bibinfo{year}{2020}\natexlab{b}.
\newblock \showarticletitle{Recovery: A multimodal repository for covid-19 news
  credibility research}. In \bibinfo{booktitle}{\emph{Proc. of CIKM}}.
\newblock


\bibitem[\protect\citeauthoryear{Zhou, Wu, and Zafarani}{Zhou
  et~al\mbox{.}}{2020c}]%
        {zhou_safe_2020}
\bibfield{author}{\bibinfo{person}{Xinyi Zhou}, \bibinfo{person}{Jindi Wu},
  {and} \bibinfo{person}{Reza Zafarani}.} \bibinfo{year}{2020}\natexlab{c}.
\newblock \showarticletitle{{SAFE}: Similarity-Aware Multi-modal Fake News
  Detection}. In \bibinfo{booktitle}{\emph{Advances in Knowledge Discovery and
  Data Mining}}. \bibinfo{publisher}{Springer International Publishing},
  \bibinfo{pages}{354--367}.
\newblock
\showISBNx{978-3-030-47436-2}


\bibitem[\protect\citeauthoryear{Zhou and Zafarani}{Zhou and Zafarani}{2018}]%
        {zhou_fake_2018}
\bibfield{author}{\bibinfo{person}{Xinyi Zhou} {and} \bibinfo{person}{Reza
  Zafarani}.} \bibinfo{year}{2018}\natexlab{}.
\newblock \showarticletitle{Fake News: A Survey of Research, Detection Methods,
  and Opportunities}.
\newblock \bibinfo{journal}{\emph{{arXiv}:1812.00315 [cs]}}
  (\bibinfo{year}{2018}).
\newblock
\showeprint[arxiv]{1812.00315}


\bibitem[\protect\citeauthoryear{Zhou and Zafarani}{Zhou and Zafarani}{2019}]%
        {zhou_network-based_2019}
\bibfield{author}{\bibinfo{person}{Xinyi Zhou} {and} \bibinfo{person}{Reza
  Zafarani}.} \bibinfo{year}{2019}\natexlab{}.
\newblock \showarticletitle{Network-based Fake News Detection: A Pattern-driven
  Approach}.
\newblock \bibinfo{journal}{\emph{{arXiv} e-prints}} (\bibinfo{year}{2019}),
  \bibinfo{pages}{arXiv:1906.04210}.
\newblock


\end{thebibliography}

\end{document}